\documentclass[english,twocolumn,transaction,10pt]{IEEEtran}
\usepackage[T1]{fontenc}
\usepackage{float}
\usepackage{amsmath}
\usepackage{amsthm}
\usepackage{amssymb}
\usepackage{stackrel}
\usepackage{graphicx}
\usepackage{setspace}
\usepackage{url}
\usepackage{color}
\usepackage{lipsum} 

\usepackage{caption}
\usepackage{subfigure}
\usepackage{stfloats}

\usepackage[ruled,linesnumbered]{algorithm2e}

\makeatletter

\theoremstyle{plain}

\theoremstyle{definition}

\theoremstyle{plain}

\theoremstyle{plain}

\IEEEoverridecommandlockouts
\usepackage{ifpdf}
\usepackage{cite}
\ifCLASSOPTIONcompsoc
\usepackage[caption=false,font=normalsize,labelfont=sf,textfont=sf]{subfig}
\else
\usepackage[caption=false,font=footnotesize]{subfig}
\fi
\usepackage{amsmath}
\usepackage{mathtools}
\usepackage{booktabs} 
\usepackage{multirow}
\usepackage{setspace}
\usepackage{lettrine} 
\usepackage{bm}

\@ifundefined{showcaptionsetup}{}{%
 \PassOptionsToPackage{caption=false}{subfig}}
\usepackage{subfig}
\makeatother

\usepackage{babel}

\newtheorem{prop}{Proposition}

\newcommand*{\QEDA}{\hfill\ensuremath{\blacksquare}} 

\begin{document}
\captionsetup[figure]{font={small}, name={Fig.}, labelsep=period}

\title{Sum-Rate Maximization for Multi-Reconfigurable Intelligent Surface-Assisted Device-to-Device Communications}
\author{Yashuai Cao,~\IEEEmembership{Student Member,~IEEE}, Tiejun Lv, \emph{Senior Member, IEEE}, Wei Ni, \emph{Senior Member, IEEE}, \\ and Zhipeng Lin, \emph{Member, IEEE}

\thanks{
Manuscript received December 27, 2020; revised April 28, 2021 and June 29, 2021; accepted August 13, 2021.
\emph{(Corresponding author: Tiejun Lv.)}

Y. Cao and T. Lv are with the School of Information and Communication Engineering, Beijing University of Posts and Telecommunications (BUPT), Beijing 100876, China (e-mail: \{yashcao, lvtiejun\}@bupt.edu.cn).

W. Ni is with Data61, Commonwealth Scientific and Industrial Research,
Sydney, NSW 2122, Australia (e-mail: wei.ni@data61.csiro.au).

Z. Lin is with the Key Laboratory of Dynamic Cognitive System of Electromagnetic Spectrum Space, College of Electronic and Information Engineering, NUAA, Nanjing 211106, China (e-mail: linlzp@ieee.org).

This paper is the extended version of its early work appeared in \cite{cao2020sum}.
}}

\maketitle
\begin{abstract}
This paper proposes to deploy multiple reconfigurable intelligent surfaces (RISs) in device-to-device (D2D)-underlaid cellular systems. The uplink sum-rate of the system is maximized by jointly optimizing the transmit powers of the users, the pairing of the cellular users (CUs) and D2D links, the receive beamforming of the base station (BS), and the configuration of the RISs, subject to the power limits and quality-of-service (QoS) of the users.
To address the non-convexity of this problem, we develop a new block coordinate descent (BCD) framework which decouples the D2D-CU pairing, power allocation and receive beamforming, from the configuration of the RISs.
Specifically, we derive closed-form expressions for the power allocation and receive beamforming under any D2D-CU pairing, which facilitates interpreting the D2D-CU pairing as a bipartite graph matching solved using the Hungarian algorithm.
We transform the configuration of the RISs into a quadratically constrained quadratic program (QCQP) with multiple quadratic constraints.
A low-complexity algorithm, named Riemannian manifold-based alternating direction method of multipliers (RM-ADMM), is developed to decompose the QCQP into simpler QCQPs with a single constraint each, and solve them efficiently in a decentralized manner.
Simulations show that the proposed algorithm can significantly improve the sum-rate of the D2D-underlaid system with a reduced complexity, as compared to its alternative based on semidefinite relaxation (SDR).
\end{abstract}


\section{Introduction}
\lettrine[lines=2]{T}{he} Internet of Things (IoT) is a promising and appealing networking paradigm, where devices can be connected in an intelligent way.
Being an integral part of the IoT \cite{8356740, 8466610, 8744244}, device-to-device (D2D) communication allows direct communication between IoT devices to improve network spectral efficiency.
Two D2D modes are typically considered in wireless networks, namely, underlaid or overlaid modes \cite{6805125, 7143335}.
In the overlaid mode, cellular users (CUs) and D2D pairs utilize orthogonal resources to avoid interference. In contrast, D2D links and CUs reuse the same spectrum to enhance spectral efficiency in a D2D-underlaid communication mode.
Severe co-channel interference could compromise the quality of service (QoS) of the CUs.
Some CUs can be far away from base station (BS) (e.g., at the cell edges) and suffer from severe propagation losses.
Their QoS is susceptible to the interference from the D2D links.
Effective resource allocation strategies have gained an upsurge of interest to suppress the interference in D2D-underlaid communications \cite{8490850, 7933260}.

A number of existing studies have attempted to improve or guarantee the QoS of both the D2D users (DUs) and the CUs \cite{8761588, 9273063}.
In \cite{6825052}, a radio resource allocation scheme was proposed for a relay-assisted D2D communication system, where the D2D pairs are far away from each other.
The authors of \cite{8385110} designed a joint beamforming and power control strategy to reduce the total transmit power consumption for both BS and DTs, while satisfying the QoS constraints for users.
The authors of \cite{7143335} guaranteed the QoS of the CUs through mode selection. The mode selection allows the DUs to switch among an underlay, overlay, or cooperative relay mode. A two-timescale resource allocation scheme was developed in \cite{8927889} to achieve a win-win situation in a D2D network, where the CUs are far away from the BS and the DUs can serve as relays to assist the CUs.
The equipment cost and self-interference of the relays were considered. However, the scheme assigns each CU with a different relay, limiting its scalability.

RIS is an emerging passive surface of engineered electromagnetic material~\cite{9133266, li2017electromagnetic, 8466374, 9020088}. A typical RIS consists of a large number of individually controllable tiles~\cite{8796365}. The properties of the RIS, such as scattering, absorption, reflection, and diffraction, can be dynamically changed by reconfiguring the phases of the tiles~\cite{9122596}. While being increasingly studied in cellular settings~\cite{9090356, 9198125, 9076830, 9039554, 8811733}, only several recent studies have attempted to integrate RISs into D2D communications.
In \cite{9301375}, an RIS-aided cell was studied in the presence of a cellular user, multiple D2D pairs, and a single RIS, where the sum-rate of the cell was maximized by optimizing the phase shifts of the RIS and the transmit power of all links.
In \cite{yang2020reconfigurable}, an RIS-assisted cellular network with D2D underlaid was studied in a quasi-static channel. Considering a single RIS, the corresponding passive beamforming was formulated as a quadratic constrained quadratic programming (QCQP) problem and solved applying off-the-shelf CVX solvers.
The authors of \cite{9322411} jointly optimized the position and the phase shifts of an RIS in D2D-underlaid systems, where the BS, CUs, and DUs all transmit a fixed power. By applying a deep Q-network (DQN) with a dynamic reward, the sum-rate of D2D and cellular networks was maximized.

By deploying RISs in D2D-underlaid cellular systems, we anticipate that the communication links can be improved for CUs and co-channel interference can be mitigated between the CUs and D2D links~\cite{yang2020reconfigurable}. Particularly, RISs are expected to improve connectivity at blind spots (i.e., the areas suffering from significant signal attenuations) by guiding the RIS-reflected paths into and out of the areas~\cite{9122596}. Consider that there could be multiple such areas in a cell, typically around the coverage boundary of the cell. The deployment of multiple RISs around the cell boundary can help improve the radio propagation and, in turn, the system throughput.

Most existing studies of RISs use the semidefinite relaxation (SDR) to solve quadratically constrained quadratic program (QCQP) problems formulated for the passive beamforming with a single constant-modulus constraint \cite{8811733, 8982186, wang2019intelligent}. Unfortunately, the SDR would incur a prohibitive complexity if multiple RISs are involved or the RIS is large, due to the fact that the number of variables grows quadratically with the number of RIS elements. Since a quadratic inequality constraint is needed for the QoS requirement of each CU, the complexity would grow rapidly with the number of CUs if the SDR is used.
The authors of \cite{8855810} proposed a low-complexity manifold optimization technique to solve the QCQP problem under a single constant-modulus constraint. However, the technique is not directly applicable in the presence of multiple inhomogeneous quadratic constraints, because the manifold optimization is only suited to constrained optimization problems that can be converted to unconstrained problems on the Riemannian manifold \cite{8125771}.


This paper presents a new approach to jointly allocating radio resources and configuring multiple distributed RISs for the uplink of a D2D-underlaid cellular system, where the RISs are deployed at the cell boundary to improve propagation. The pairing of CUs and D2D links (to reuse the subchannels allocated to the CUs), the transmit powers of the users, the receive beamformers of the BS, and the passive beamformers of the RISs are jointly optimized to maximize the sum-rate of the system while guaranteeing the minimum data rate requirements (or the QoS) of the CUs.
The joint optimization is non-trivial, because of coupling between the passive beamformers of the RISs and other wireless control variables, constant-modulus phase shifts of the RISs, and the discrete nature of D2D-CU pairing. The joint optimization is a mixed-integer non-linear problem (MINLP), which is non-convex and NP-hard.
It is even more challenging due to the consideration of the QoS of the CUs. This is because a quadratic inequality constraint is needed to account for the QoS requirement of every CU~\cite{8385110}, leading to a rapid growth of complexity with the increase of CUs.


The key contributions of this paper are summarized as follows.
\begin{itemize}
\item Given the non-convexity of the considered problem, we propose a new block coordinate descent (BCD)-based framework, which decouples the problem between D2D-CU matching and the passive beamforming of the RISs.
\item We derive the closed-form expressions for the optimal transmit powers of the users and receive beamforming of the BS under any D2D-CU matching, interpret the D2D-CU matching as a bipartite graph (BG) accordingly, and solve the BG matching using the Hungarian algorithm.
\item We exploit quadratic transform techniques to reformulate the passive beamforming of the RISs into a non-convex and inhomogeneous QCQP, which can be readily solved using the classic SDR technique.
\item A low-complexity algorithm, named Riemannian manifold-based alternating direction method of multipliers (RM-ADMM), is developed to efficiently solve the QCQP, where quadratic constraints are decoupled into simpler QCQPs to be pursued in parallel.
\end{itemize}
Extensive simulations verify the superiority of the proposed algorithm to its state-of-the-art SDR-based alternative in terms of sum-rate and efficiency. The role of the RISs in the considered system is investigated with useful insights drawn.

The rest of this paper is organized as follows. In Section \ref{section:system}, we present the system model and problem formulation. We analyze and reformulate the problem in Section \ref{section:method}, followed by elaborating on the proposed algorithm in Section \ref{section:props}. Section \ref{section:sim} presents simulation results, followed by conclusions in Section \ref{section:con}.

\emph{Notations}: Lower-case boldface denotes column vectors, and upper-case boldface denotes matrices; $(\cdot)^{\ast}$, $(\cdot)^{-1}$, $(\cdot)^{\mathsf{T}}$, and $(\cdot)^{\mathsf{H}}$ denote the conjugate, matrix inversion, transpose, and conjugate transpose, respectively; $\mathsf{Diag}\{\mathbf{a}\}$ returns the matrix with $\mathbf{a}$ on its diagonal; $\mathsf{vec}(\mathbf{A})$ denotes the vectorization of a matrix $\mathbf{A}$; $\mathrm{Re}\{\cdot\}$ indicates the real part of complex values; $\mathbf{I}_{A}$ is the $A\times A$ identity; and $\circ$ denotes the Hadamard product operator.

\section{System Model and Problem Formulation}\label{section:system}
\subsection{System Model}
We consider the uplink of a D2D-underlaid single-input multiple-output (SIMO) cellular network, where there is an $M$-antenna BS, $K$ single-antenna CUs, and $J$ pairs of single-antenna DUs.
The CUs and D2D pairs are collected by $\mathcal{C}=\{1,2,\cdots,K\}$ and $\mathcal{D}=\{1,2,\cdots,J\}$, respectively.
Assume that the system has $K$ orthogonal subchannels. Every subchannel is preassigned to a CU. A D2D pair can reuse one of the subchannels.
Also assume that $K\geq J$, so that a D2D pair can be allocated with a subchannel and the transmit rate of the D2D pair is always non-zero.
Let $\rho_{jk}$ denote the reuse indicator. $\rho_{jk}=1$ if the $j$-th D2D pair reuses the subchannel of the $k$-th CU. $\rho_{jk}=0$, otherwise. $\bm{\rho}=[\rho_{11},\rho_{12},\cdots,\rho_{JK}]^{\mathsf{T}}$ collects all the reuse indicators.
There is co-channel interference between the CU and the D2D pair reusing the same subchannel. Assume that the CUs are around the cell edge. $L$ RISs are deployed at the cell edge to improve radio propagation. Each RIS comprises $N$ reflecting elements. Let $\mathcal{L}=\{1,2,\cdots,L\}$ collect the indexes to RISs.

\begin{figure}[t]
	\centering{}\includegraphics[scale=0.4]{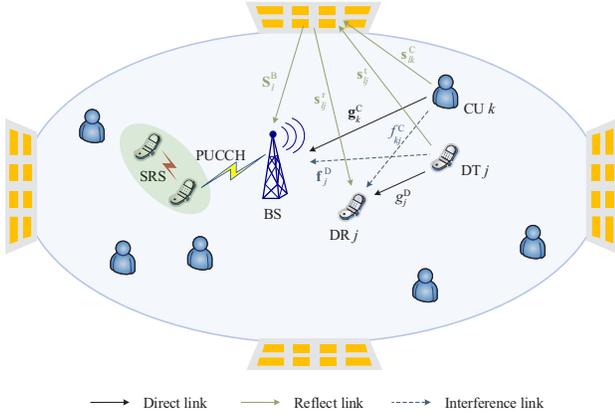}
	\caption{RIS assisted D2D-underlaid cellular system.}
	\label{fig:system}
\end{figure}

As shown in Fig. \ref{fig:system}, the channels from the $k$-th CU to the BS and from the $j$-th D2D transmitter (DT) to the $j$-th D2D receiver (DR) are denoted by $\mathbf g_{k}^{\mathrm{C}} \in \mathbb{C}^{M \times 1}$ and $g_{j}^{\mathrm{D}} \in \mathbb{C}$, respectively; the interference channels from the $k$-th CU to the $j$-th DR and from the $j$-th DT to the BS are denoted by $f_{kj}^{\mathrm{C}} \in \mathbb{C}$ and $\mathbf f_{j}^{\mathrm{D}} \in \mathbb{C}^{M \times 1}$, respectively; the channels from the $k$-th CU to the $l$-th RIS, from the $l$-th RIS to the BS, from the $j$-th DT to the $l$-th RIS and from the $l$-th RIS to the $j$-th DR are denoted by $\mathbf s_{lk}^{\mathrm{C}} \in \mathbb{C}^{N \times 1}$, $\mathbf S_{l}^{\mathrm{B}} \in \mathbb{C}^{M \times N}$, $\mathbf s_{lj}^{\mathrm{t}} \in \mathbb{C}^{N \times 1}$ and $\mathbf s_{lj}^{\mathrm{r}} \in \mathbb{C}^{N \times 1}$, respectively.
As specified in 3GPP LTE standards~\cite{etsi136lte}, all users, including those forming D2D links, feed their CSI back to the BS via the physical uplink control channel (PUCCH)~\cite{8950342}. Each user can access the sounding reference signal (SRS) channel~\cite{dahlman20134g} regularly for CSI estimation~\cite{7491359}.
Several methods have been developed to estimate the RIS-assisted cascaded channels, such as an ON/OFF protocol-based method~\cite{8683663}, matrix-calibration-based method~\cite{9133156}, and discrete Fourier transform (DFT) protocol-based method~\cite{9087848}.
Particularly, the DFT protocol-based method~\cite{9087848} is optimal in terms of channel estimation accuracy~\cite{9328501}, where the phase shifts of the RIS elements are set to be a column vector of the DFT matrix in each subframe and the minimum mean squared error (MMSE) estimator is used to estimate jointly the channels arriving at and departing from an RIS. The estimation of the RIS-aided channels is beyond the scope of this paper.

Let $\mathbf\Theta_l=\mathsf{Diag}\{ e^{j \theta_{l1}},e^{j \theta_{l2}},\cdots, e^{j \theta_{lN}} \}$ be the phase shift matrix of the $l$-th RIS. $\mathcal{S}_{\Theta}=\{\mathbf\Theta_l\}_{l=1}^{L}$ is the set of phase shift matrices.
The phase shifts of the RISs and the transmit powers of the CUs and DTs are optimized to mitigate the co-channel interference and maximize the uplink sum-rate.
The optimization is conducted centrally at the BS. One reason is that the passive RISs do not actively transmit or receive signals. Therefore, the use of RISs would typically involve centralized coordination (often via wired links) to control the phase shifts of RIS tiles, e.g., by configuring the voltages of the tiles~\cite{8796365, 9122596, 9129476}. Another reason is that, being part of a cellular network, D2D links reusing a cellular spectrum need to be managed by the network~\cite{7728080} to maintain the acceptable quality of paid services for cellular users.
As a matter of fact, the setup of D2D links needs to be granted and coordinated centrally by the BSs in 3GPP standards~\cite{6231164, 3gpp2012, 6807945}.

\subsection{Problem Formulation}
Let $P_{j}^{\mathrm{D}}$ and $P_{k}^{\mathrm{C}}$ denote the transmit powers of the $j$-th DT and the $k$-th CU, respectively. $\mathcal{S}_P=\{\mathbf{p}^{\mathrm{D}},\mathbf{p}^{\mathrm{C}}\}$ collects all the allocated powers, where $\mathbf{p}^{\mathrm{D}}=[P_{1}^{\mathrm{D}},\cdots,P_{J}^{\mathrm{D}}]^{\mathsf{T}}$ and $\mathbf{p}^{\mathrm{C}}=[P_{1}^{\mathrm{C}},\cdots,P_{K}^{\mathrm{C}}]^{\mathsf{T}}$.
Suppose that the $j$-th D2D link and the $k$-th CU are paired to reuse the subchannel preassigned to the CU, i.e., $\rho_{jk}=1$. As assumed in Section II-A, $K\geq J$ and therefore, for any $j$, there must exist such $k$ that $\rho_{jk}=1$.
The signal-to-interference-plus-noise ratio (SINR) at DR $j$ is given by
\begin{align}
\gamma_{j}^{\mathrm{D}} = \frac{P_{j}^{\mathrm{D}} \vert g_{j}^{\mathrm{D}} + \sum\limits_{l\in \mathcal{L}} {\mathbf {s}_{lj}^{\mathrm{r}}}^{\mathsf H}\mathbf\Theta_l\mathbf{s}_{lj}^{\mathrm{t}} \vert^2}{ \sum\limits_{k'\in \mathcal{C}} \rho_{jk'} P_{k'}^{\mathrm{C}} \vert f_{k'j}^{\mathrm{C}} + \sum\limits_{l\in \mathcal{L}}{\mathbf s_{lj}^{\mathrm{r}}}^{\mathsf H}\mathbf\Theta_l\mathbf s_{lk'}^{\mathrm{C}} \vert^2 + \sigma_d^2}, \label{eq:gamma_d}
\end{align}
where $\sigma_D^2$ is the noise power at the DR.
The SINR of the uplink link from CU $k$ to the BS is given by
\begin{align}
\gamma_{k}^{\mathrm{C}} = \frac{P_{k}^{\mathrm{C}} \vert \mathbf w_{k}^{\mathsf H} (\mathbf g_{k}^{\mathrm{C}} + \sum\limits_{l\in \mathcal{L}}\mathbf S_{l}^{\mathrm{B}}\mathbf\Theta_l\mathbf s_{lk}^{\mathrm{C}}) \vert^2}{\sum\limits_{j' \in\mathcal{D}} \rho_{j' k} P_{j'}^{\mathrm{D}} \vert \mathbf w_{k}^{\mathsf H} ( \mathbf f_{j'}^{\mathrm{D}} + \sum\limits_{l\in \mathcal{L}}\mathbf S_{l}^{\mathrm{B}}\mathbf\Theta_l\mathbf s_{lj'}^{\mathrm{t}}) \vert^2 + \sigma_b^2}, \label{eq:gamma_c}
\end{align}
where $\sigma_b^2$ is the noise power at the BS, and $\mathbf w_k \in \mathbb C^{M \times 1}$ is the unit-norm receive beamforming vector of the BS for CU $k$.
Then, the data rate of the $j$-th D2D pair is $R_{j}^{\mathrm{D}}=\log(1+\gamma_{j}^{\mathrm{D}})$ and the data rate of the $k$-th CU is $R_{k}^{\mathrm{C}}=\log(1+\gamma_{k}^{\mathrm{C}})$.

We aim to maximize the sum-rate of the system, subject to the power limits of all users and the SINR requirements of the CUs, as given by
\begin{subequations}
\begin{align}
\text{(P1)}:  &\ \underset{\{\bm{\rho}, \mathbf{p}^{\mathrm{C}}, \mathbf{p}^{\mathrm{D}}, \mathbf{W}, \mathcal{S}_{\Theta}\}}{\max} \  R(\bm{\rho}, \mathbf{p}^{\mathrm{C}}, \mathbf{p}^{\mathrm{D}}, \mathbf{W}, \mathcal{S}_{\Theta}) \nonumber  \\
\mathrm{s.t.}
\quad &  0 < P_{j}^{\mathrm{D}} \le P_{\mathrm{D}}^{\max}, \ \forall j \in \mathcal{D} ,\label{eq:pd} \\
& 0 < P_{k}^{\mathrm{C}} \le P_{\mathrm{C}}^{\max}, \ \forall k\in \mathcal{C},\label{eq:pc} \\
& \gamma_{k}^{\mathrm{C}} \ge \gamma_{\mathrm{C}}^{\mathrm{th}}, \ \forall k\in \mathcal{C},\label{eq:qos_c} \\
& \Vert \mathbf w_k \Vert^2 = 1, \ \forall k\in \mathcal{C}, \label{eq:re_bf}  \\
& \rho_{jk} \in \{0,1\}, \ \forall j \in \mathcal{D},\ \forall k\in \mathcal{C},  \label{eq:resue_1} \\
& \sum_{k \in \mathcal{C}}\rho_{jk} = 1, \ \forall j \in \mathcal{D}, \label{eq:resue_2}\\
& \sum_{j \in \mathcal{D}}\rho_{jk} \le 1, \ \forall k\in \mathcal{C}, \label{eq:resue_3} \\
& \theta_{ln} \in [ 0, 2\pi ), \ \forall l,n, \label{eq:phase}
\end{align}
\end{subequations}
where $\mathbf{W}=[\mathbf{w}_1,\cdots, \mathbf{w}_K]$; $R=\sum_{j\in\mathcal{D}} R_{j}^{\mathrm{D}} + \sum_{k\in\mathcal{C}} R_{k}^{\mathrm{C}}$; $P_{\mathrm{C}}^{\max}$ and $P_{\mathrm{D}}^{\max}$ are the maximum transmit powers of the CUs and DTs, respectively; $\gamma_{\mathrm{C}}^{\mathrm{th}}$ is the minimum SINR requirement of the CUs to guarantee the QoS of the CUs\footnote{Since strong transmission always exists among D2D links \cite{7277112, 6953070, 7417275}, in this paper we focus  on guaranteeing the QoS of CUs.}. (\ref{eq:re_bf}) is the unit-norm constraint of the receive beamforming;
(\ref{eq:resue_1}) enforces the binary nature of the reuse indicator;
(\ref{eq:resue_2}) specifies that each D2D link can only match with a CU; (\ref{eq:resue_3}) indicates that each CU can be matched to at most a D2D link;
and (\ref{eq:phase}) specifies the range of the phase shifts of the RIS.

\section{Proposed Alternating Optimization}\label{section:method}
Problem (P1) is a mixed integer programming problem, where the passive beamforming of the RISs is closely coupled with the D2D-CU matching, power allocation and receive beamforming. The objective of the problem is a sum-of-logarithms. For this reason, Problem (P1) is non-convex and difficult to solve. We propose the new BCD framework that decouples and optimizes the optimization variables in an alternating manner. By using the framework, Problem (P1) is decomposed into two subproblems, namely, 1) D2D-CU pairing, the power allocation and receive beamforming, given fixed configuration of the RISs; and 2) configuration of the RISs, given fixed D2D-CU pairing, power allocation, and receive beamforming.

\subsection{D2D-CU Pairing, Power Allocation and Receive Beamforming}
We first match the D2D-CU pairs given fixed the configuration of the RISs $\mathbf{\Theta}_l$.
For notational brevity, we denote $h_{j}^{\mathrm{D}}\triangleq g_{j}^{\mathrm{D}} + \sum_{l\in \mathcal{L}}{\mathbf{s}_{lj}^{\mathrm{r}}}^{\mathsf H}\mathbf{\Theta}_l\mathbf{s}_{lj}^{\mathrm{t}}$, $h_{kj}^{\mathrm{C}}\triangleq f_{kj}^{\mathrm{C}} + \sum_{l\in \mathcal{L}}{\mathbf{s}_{lj}^{\mathrm{r}}}^{\mathsf H}\mathbf{\Theta}_l\mathbf{s}_{lk}^{\mathrm{C}}$, $\mathbf{h}_{k}^{\mathrm{C}}\triangleq\mathbf{g}_{k}^{\mathrm{C}}+\sum_{l\in \mathcal{L}}\mathbf{S}_{l}^{\mathrm{B}}\mathbf{\Theta}_l\mathbf{s}_{lk}^{\mathrm{C}}$, and $\mathbf h_{j}^{\mathrm{D}}\triangleq\mathbf{f}_{j}^{\mathrm{D}}+\sum_{l\in \mathcal{L}}\mathbf{S}_{l}^{\mathrm{B}}\mathbf{\Theta}_l\mathbf{s}_{lj}^{\mathrm{t}}$.
The coefficients $h_{j}^{\mathrm{D}}$, $h_{kj}^{\mathrm{C}}$, $\mathbf{h}_{k}^{\mathrm{C}}$ and $\mathbf{h}_{j}^{\mathrm{D}}$ in (\ref{eq:gamma_d}) and (\ref{eq:gamma_c}) are fixed.

The D2D-CU matching is interpreted as a bipartite graph (BG) between the CUs and the D2D pairs.
We create a BG between CUs and D2D pairs, denoted by $\mathcal{G}_0(\mathcal{V}_0, \mathcal{E}_0)$, where $\mathcal{V}_0$ collects the vertexes and $\mathcal{E}_0$ collects the edges. $\mathcal{V}_0$ is further divided into two disjoint subsets, $\mathcal{C}$ and $\mathcal{D}$, to collect the CUs and DUs, respectively.
The edge between D2D pair $j$ and CU $k$ indicates that the D2D pair reuses the resource of the CU. The weight of the edge is the maximum achievable sum-rate by optimizing the transmit powers and receive beamformers in the subchannel allocated to CU $k$, i.e., $R_{j,k}^{\mathrm{opt}}=R_{j}^{\mathrm{D}} +R_{k}^{\mathrm{C}} \vert_{ P_{k}^{\mathrm{C,opt}}, P_{j}^{\mathrm{D,opt}},\mathbf{w}_{k}^{\mathrm{opt}} }$.


When considering the possible pairing between the $j$-th D2D pair and the $k$-th CU, i.e., D2D-CU pair $(j,k)$, we have $\rho_{jk}=1$.
The maximum sum-rate of the D2D-CU pair $(j,k)$ can be obtained by solving
\begin{align}
\text{(P2)}:  \underset{P^{\mathrm{C}}_k, P^{\mathrm{D}}_j, \mathbf{w}_k}{\max}  & R^{\mathrm{D}}_j(P^{\mathrm{C}}_k, P^{\mathrm{D}}_j)+R^{\mathrm{C}}_k(P^{\mathrm{C}}_k, P^{\mathrm{D}}_j,\mathbf{w}_k) \nonumber  \\
\mathrm{s.t.}
\quad &  \text{(\ref{eq:pd})}-\text{(\ref{eq:re_bf})}. \nonumber
\end{align}

We can obtain the optimal solution to $\mathbf{w}_k$ in Problem (P2) by maximizing $\gamma_k^{\mathrm{C}}$ in (\ref{eq:gamma_c}).
By employing the Rayleigh quotient maximization which is optimal \cite{7151842}, we have
\begin{align}
\mathbf{w}_k (P^{\mathrm{D}}_j) &= \underset{\mathbf{w}_k: \|\mathbf{w}_k^{2}\|{=}1}{\arg\max} \quad \frac {P^{C}_k \mathbf{w}_k^{\mathsf H} {\mathbf{h}}^{\mathrm{C}}_k {\mathbf{h}}_k^{\mathrm{C}^{\mathsf H}}\mathbf{w}_k}{\mathbf{w}_k^{\mathsf H} \left( P^{\mathrm{D}}_j  {\mathbf{h}}_j^{\mathrm{D}} {\mathbf{h}}_j^{{\mathrm{D}}^{\mathsf H}}  + \sigma_b^2\mathbf{I}_M \right)\mathbf{w}_k} \nonumber\\
&=\frac{  (P^{\mathrm{D}}_j \mathbf{h}_j^{\mathrm{D}} {\mathbf{h}_j^{\mathrm{D}}}^{\mathsf H}  + \sigma_b^2 \mathbf{I}_M )^{-1} \mathbf{h}_k^{\mathrm{C}} }{\Vert (P^{\mathrm{D}}_j \mathbf{h}^{\mathrm{D}}_j {\mathbf{h}_j^{\mathrm{D}}}^{\mathsf H}  + \sigma_b^2 \mathbf{I}_M )^{-1} \mathbf{h}_k^{\mathrm{C}} \Vert}. \label{eq:opt_w}
\end{align}

\begin{figure}[t]
    \centering{}\includegraphics[scale=0.35]{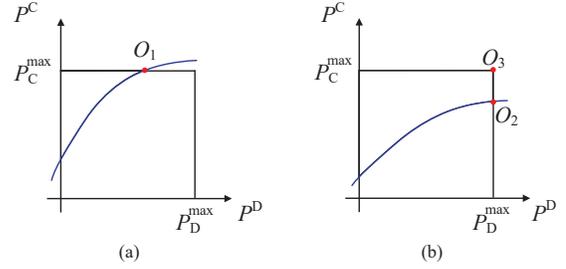}
	\caption{The feasible region in the $P^{\mathrm{D}}$-$P^{\mathrm{C}}$ power plane.}
	\label{fig:region}
\end{figure}

Next, we investigate the power allocation problem with the optimal $\mathbf{w}_k$.
Plugging (\ref{eq:opt_w}) into (\ref{eq:gamma_c}), we can reformulate the SINR constraint (\ref{eq:qos_c}) as
\begin{align}
P^{\mathrm{C}}_k & \ge \tilde{\gamma}_{\mathrm{C}} \left( 1- \frac{\lambda_{1} P^{\mathrm{D}}_j }{P^{\mathrm{D}}_j + \lambda_{2}} \right)^{-1}, \label{eq:line_c}
\end{align}
where $\tilde{\gamma}_{\mathrm{C}}=\frac{\sigma_b^2 \gamma_{\mathrm{C}}^{\mathrm{th}}}{\Vert \mathbf{h}^{\mathrm{C}}_k \Vert^2}$, $\lambda_{1} = (\frac{\vert {\mathbf{h}^{\mathrm{C}}_k}^{\mathsf H} \mathbf{h}^{\mathrm{D}}_j \vert}{\Vert \mathbf{h}_k^{\mathrm{C}}\Vert \cdot \Vert\mathbf{h}_j^{\mathrm{D}} \Vert} ) ^2 \in [0, 1]$, and $\lambda_{2}=\frac{\sigma_b^2}{\Vert \mathbf{h}_j^{\mathrm{D}} \Vert^2}$ for notational brevity.
To characterize the feasible region on the $P^{\mathrm{D}}_j$-$P^{\mathrm{C}}_k$ power plane, we take equality in constraint (\ref{eq:line_c}) and obtain a concave increasing function of $P^{\mathrm{D}}_j$, i.e., $P^{\mathrm{C}}_k(P^{\mathrm{D}}_j) = \tilde{\gamma}_{\mathrm{C}} \left( 1- \frac{\lambda_{1} P^{\mathrm{D}}_j}{P^{\mathrm{D}}_j + \lambda_{2}} \right)^{-1}$, as plotted in Fig. \ref{fig:region}. Let $\mathcal P$ be the feasible solution region of the transmit powers of the CU and DT, which is the green shaded area in Fig. \ref{fig:region}.

\begin{prop}
If the feasible region $\mathcal P$ is not empty, the optimal transmit powers $(\hat{P}^{\mathrm{D}}_j, \hat{P}^{\mathrm{C}}_k)$ are in the candidate set $\{O_1, O_2, O_3\}$.
\end{prop}

\begin{IEEEproof}
See Appendix \ref{prf1}.
\end{IEEEproof}

According to Proposition 1, we can obtain closed-form expressions for the optimal transmit powers in the following three cases:
\begin{itemize}
\item According to (\ref{eq:line_c}), the D2D pair cannot reuse the subchannel of the CU if $\tilde{\gamma}_{\mathrm{C}} > {P}_{\mathrm{C}}^{\max}$, since there are no feasible points. In this case, we have $\rho_{jk}=0$ and the maximum rate of CU can be achieved by setting $P^{\mathrm{C}}_k={P}_{\mathrm{C}}^{\max}$ and $\mathbf{w}_k = \frac{\mathbf{h}_k^{\mathrm{C}} }{\Vert \mathbf{h}_k^{\mathrm{C}} \Vert }$.
\item If ${P}_{\mathrm{C}}^{\max}<I_{\mathrm{C}}$, as shown in Fig. \ref{fig:region}(a), $O_1=(\frac{\lambda_{2}(\tilde{\gamma}_{\mathrm{C}} - P_{\mathrm{C}}^{\max})}{(1-\lambda_{1}) P_{\mathrm{C}}^{\max} - \tilde{\gamma}_{\mathrm{C}}},P_{\mathrm{C}}^{\max})$ is the optimal power solution.
\item If $I_{\mathrm{C}} \le {P}_{\mathrm{C}}^{\max}$, as shown in Fig. \ref{fig:region}(b), $(\hat{P}^{\mathrm{D}}_j, \hat{P}^{\mathrm{C}}_k)$ is chosen from the candidate set $\{O_2, O_3\}$, where $O_2=(P_{\mathrm{D}}^{\max}, I_{\mathrm{C}})$ and $O_3=({P}_{\mathrm{D}}^{\max}, {P}_{\mathrm{C}}^{\max})$.
\end{itemize}
Here, $I_{\mathrm{C}} = \frac{\tilde{\gamma}_{\mathrm{C}} ({P}_{\mathrm{D}}^{\max} + \lambda_{2})}{(1-\lambda_{1}) {P}_{\mathrm{D}}^{\max} + \lambda_{2}}$ is the ordinate of the intersection point $O_2$. Once $(\hat{P}^{\mathrm{D}}_j, \hat{P}^{\mathrm{C}}_k)$ is determined, the optimal receive beamforming vector $\mathbf{w}^{\mathrm{opt}}$ is obtained accordingly by (\ref{eq:opt_w}).
The SINR constraints (3c) is satisfied as long as Problem (P2) is feasible. With the weights of all possible D2D-CU pairs, i.e., $R_{j,k}^{\mathrm{opt}},\forall j,k$, obtained by solving problem (P2), the D2D-CU matching problem can be cast as
\begin{align}
\text{(P3)}:  &\ \underset{\{ \bm{\rho} \}}{\max} \  \sum_{k\in\mathcal{C}} \sum_{j\in\mathcal{D}} \rho_{jk} R_{j,k}^{\mathrm{opt}} + \sum_{k\in\mathcal{C}} (1-\sum_{j\in\mathcal{D}} \rho_{jk} ) R_{k}^{\mathrm{C, opt}}   \nonumber  \\
\mathrm{s.t.}
& \quad  \text{(\ref{eq:resue_1})}-\text{(\ref{eq:resue_3})}. \nonumber
\end{align}
Problem (P3) is a standard maximum weighted bipartite matching, because of the binary constraints resulting from the user pairing.
According to graph theory, the D2D-CU matching problem is a maximum weighted bipartite matching problem,
and can be efficiently solved using the Hungarian algorithm,
which is a celebrated combinational optimization algorithm and solves assignment problems, e.g., Problem (P3) in polynomial time \cite{jungnickel2008}.

\subsection{Problem Reformulation for Passive Phase-Shift Design}
Given $\{\rho_{jk}, P^{\mathrm{C}}_{k}, P^{\mathrm{D}}_{j}, \mathbf{w}_k\}$, we optimize the passive beamforming of the RISs. For ease of illustration, let $\tilde{g}^{\mathrm{D}}_{j}\triangleq\sqrt{P^{\mathrm{D}}_{j}} g^{\mathrm{D}}_{j}$, $\mathbf{a}_{lj}^{\mathsf{H}}\triangleq\sqrt{P^{\mathrm{D}}_{j}}{\mathbf{s}^{\mathrm{r}}_{lj}}^{\mathsf{H}} \mathsf{Diag}\{\mathbf{s}^{\mathrm{t}}_{lj}\}$, $\tilde{f}^{\mathrm{C}}_{kj}\triangleq\sqrt{P^{\mathrm{C}}_{k}} f^{\mathrm{C}}_{kj}$, $\mathbf{b}_{lkj}^{\mathsf{H}}\triangleq\sqrt{P^{\mathrm{C}}_{k}}{\mathbf{s}^{\mathrm{r}}_{lj}}^{\mathsf{H}}\mathsf{Diag}\{\mathbf{s}^{\mathrm{C}}_{lk}\}$, $\tilde{g}^{\mathrm{C}}_{k}\triangleq\sqrt{P^{\mathrm{C}}_{k}}\mathbf{w}_k^{\mathsf{H}} \mathbf{g}^{\mathrm{C}}_{k}$, $\bm{\alpha}_{lk}^{\mathsf{H}}=\sqrt{P^{\mathrm{C}}_{k}}\mathbf{w}_k^{\mathsf{H}}\mathbf{S}^{\mathrm{B}}_{l} \mathsf{Diag}\{\mathbf{s}^{\mathrm{C}}_{lk}\}$, $\tilde{f}^{\mathrm{D}}_{j}\triangleq\sqrt{P^{\mathrm{D}}_{j}}\mathbf{w}_k^{\mathsf{H}} \mathbf{f}^{\mathrm{D}}_{j}$, $\bm{\beta}_{lj}^{\mathsf{H}}=\sqrt{P^{\mathrm{D}}_{j}}\mathbf{w}_k^{\mathsf{H}}\mathbf{S}^{\mathrm{B}}_{l} \mathsf{Diag}\{\mathbf{s}^{\mathrm{t}}_{lj} \}$, and $\bm{\theta}_l\triangleq[e^{j \theta_{l,1}},e^{j \theta_{l,2}},\cdots, e^{j \theta_{l,N}}]^{\mathsf T}$.
Then, (\ref{eq:gamma_d}) and (\ref{eq:gamma_c}) are rewritten as
\small
\begin{align}
\gamma^{\mathrm{D}}_{j}
&= \frac{ {P^{\mathrm{D}}_{j}} \vert g^{\mathrm{D}}_{j} + \sum_{l=1}^{L} {\mathbf{s}^{\mathrm{r}}_{lj}}^{\mathsf H} \mathsf{Diag}\{\mathbf{s}^{\mathrm{t}}_{lj}\} \bm{\theta}_l \vert^2}{ {\sum_{k=1}^{K}\rho_{jk} P^{\mathrm{C}}_{k}} \vert f^{\mathrm{C}}_{kj} + \sum_{l=1}^{L} {\mathbf{s}^{\mathrm{r}}_{lj}}^{\mathsf{H}}\mathsf{Diag}\{\mathbf{s}^{\mathrm{C}}_{lk}\}\bm{\theta}_l\vert^2 + \sigma_{d}^2}, \nonumber\\
&= \frac{ \vert \tilde{g}^{\mathrm{D}}_{j} + \sum_{l=1}^{L} \mathbf{a}_{lj}^{\mathsf{H}} \bm{\theta}_l \vert^2}{ \sum_{k=1}^{K}\rho_{jk} \vert \tilde{f}^{\mathrm{C}}_{kj} + \sum_{l=1}^{L} \mathbf{b}_{lkj}^{\mathsf{H}} \bm{\theta}_l \vert^2 + \sigma_d^2}, \label{eq:th_gd} \\
\gamma^{\mathrm{C}}_{k}
&= \frac{ {P^{\mathrm{C}}_{k}}\vert \mathbf{w}_{k}^{\mathsf{H}} \mathbf{g}^{\mathrm{C}}_{k}+ \sum_{l=1}^{L}\mathbf{w}_{k}^{\mathsf{H}} \mathbf{S}^{\mathrm{B}}_{l} \mathsf{Diag}\{\mathbf{s}^{\mathrm{C}}_{lk}\} \bm{\theta}_l  \vert^2}{ \sum_{j=1}^{J}\rho_{jk} {P^{\mathrm{D}}_{j}}\vert \mathbf{w}_{k}^{\mathsf{H}} \mathbf{f}^{\mathrm{D}}_{j} + \sum_{l=1}^{L} \mathbf{w}_{k}^{\mathsf{H}} \mathbf{S}^{\mathrm{B}}_{l} \mathsf{Diag}\{\mathbf{s}^{\mathrm{t}}_{lj} \} \bm{\theta}_l \vert^2 + \sigma_b^2}, \nonumber\\
&= \frac{ \vert \tilde{g}^{\mathrm{C}}_{k} + \sum_{l=1}^{L}\bm{\alpha}_{lk}^{\mathsf{H}} \bm{\theta}_l \vert^2}{ \sum_{j=1}^{J}\rho_{jk} \vert \tilde{f}^{\mathrm{D}}_{j} + \sum_{l=1}^{L}\bm{\beta}_{lj}^{\mathsf{H}} \bm{\theta}_l \vert^2 + \sigma_b^2}, \label{eq:th_gc}
\end{align}
\normalsize
By letting $\mathbf{a}_{j} = [\mathbf{a}_{1,j}^{\mathsf{T}},\cdots, \mathbf{a}_{L,j}^{\mathsf{T}}]^{\mathsf{T}}$, $\mathbf{b}_{kj}=[\mathbf{b}_{1kj}^{\mathsf{T}},\cdots, \mathbf{b}_{Lkj}]^{\mathsf{T}}$, $\bm{\alpha}_{k}=[\bm{\alpha}_{1k}^{\mathsf{T}},\cdots,\bm{\alpha}_{Lk}^{\mathsf{T}}]^{\mathsf{T}}$, $\bm{\beta}_{j}=[\bm{\beta}_{1j}^{\mathsf{T}},\cdots,\bm{\beta}_{Lj}^{\mathsf{T}}]^{\mathsf{T}}$ and $\bm{\phi}=[\bm{\theta}_1^{\mathsf{T}},\cdots, \bm{\theta}_L^{\mathsf{T}}]^{\mathsf{T}}$, we can rewrite (\ref{eq:th_gd}) and (\ref{eq:th_gc}) as
\begin{align}
\gamma^{\mathrm{D}}_{j}=\frac{ A^{d}_{j}(\bm{\phi}) }{ \sum_{k=1}^{K} \rho_{jk} B^{c}_{kj}(\bm{\phi}) + \sigma_d^2}, \label{eq:th_gd_re} \\
\gamma^{\mathrm{C}}_{k}=\frac{ A^{c}_{k}(\bm{\phi}) }{ \sum_{j=1}^{J} \rho_{jk} B^{d}_{j}(\bm{\phi}) + \sigma_b^2}, \label{eq:th_gc_re}
\end{align}
where $A^{d}_{j}(\bm{\phi}) = \vert \tilde{g}^{\mathrm{D}}_{j} + \mathbf{a}_{j}^{\mathsf H}\bm{\phi} \vert^2$, $B^{c}_{kj}(\bm{\phi}) = \vert \tilde{f}^{\mathrm{C}}_{kj} + \mathbf{b}_{kj}^{\mathsf H}\bm{\phi} \vert^2$, $A^{c}_{k}(\bm{\phi}) = \vert \tilde{g}^{\mathrm{C}}_{k} + \bm{\alpha}_{k}^{\mathsf H}\bm{\phi} \vert^2$, and $B^{d}_{j}(\bm{\phi}) = \vert \tilde{f}^{\mathrm{D}}_{j} + \bm{\beta}_{j}^{\mathsf H}\bm{\phi} \vert^2$.
Problem (P1) is reduced to
\begin{subequations}
\begin{align}
\text{(P4)}: \quad&  \underset{ \bm{\phi} }{\max} \quad \sum_{j=1}^{J} R^{\mathrm{D}}_{j}(\bm{\phi}) + \sum_{k=1}^{K} R^{\mathrm{C}}_{k}(\bm{\phi}) \nonumber  \\
\mathrm{s.t.} \quad
& \gamma_{\mathrm{C}}^{\mathrm{th}} \sum_{j=1}^{J}\rho_{jk} ({ B^{d}_{j}(\bm{\phi}) + \sigma_b^2}) - { A^{c}_{k}(\bm{\phi})}
\le 0, \ \forall k,  \label{eq:gc_cond} \\
& \vert \theta_{ln} \vert = 1, \ \forall l, \forall n. \label{eq:phase_cond}
\end{align}
\end{subequations}

\begin{prop}
Problem (P4) is equivalent to
\begin{align}
\text{(P5)}:  \underset{ \bm{\phi},\{\zeta^{\mathrm{D}}_{j},\zeta^{\mathrm{C}}_{k}\} }{\max} \ & \sum_{j=1}^{J} \digamma(\zeta^{\mathrm{D}}_{j},\gamma^{\mathrm{D}}_{j}) + \sum_{k=1}^{K}\digamma(\zeta^{\mathrm{C}}_{k},\gamma^{\mathrm{C}}_{k}) \nonumber  \\
\mathrm{s.t.} \quad
& \text{(\ref{eq:gc_cond})}, \ \text{(\ref{eq:phase_cond})}, \nonumber
\end{align}
where $\digamma(\zeta,\gamma)=\log(1+\zeta)-\zeta+\frac{(1+\zeta)\gamma}{1+\gamma}$; and $\zeta^{\mathrm{D}}_{j}$ and $\zeta^{\mathrm{C}}_{k}$ are auxiliary variables associated with $\gamma^{\mathrm{D}}_{j}$ and $\gamma^{\mathrm{C}}_{k}$, respectively.
Given $\{\gamma^{\mathrm{D}}_{j},\gamma^{\mathrm{C}}_{k}\}$, the optimal $\zeta^{\mathrm{D}}_{j}$ is equal to $\gamma^{\mathrm{D}}_{j}$ and the optimal $\zeta^{\mathrm{C}}_{k}$ is equal to $\gamma^{\mathrm{C}}_{k}$.
\end{prop}

\begin{IEEEproof}
See Appendix \ref{prf2}.
\end{IEEEproof}

For given $\{\zeta^{\mathrm{D}}_{j}, \zeta^{\mathrm{C}}_{k}\}$, optimizing $\bm{\phi}$ in (P5) becomes a multiple-ratio fractional programming (MRFP) problem, as given by
\begin{align}
\text{(P6)}: \  \underset{ \bm{\phi} }{\max} \quad & \tilde{\digamma}(\bm{\phi}) = \sum_{j=1}^{J} \frac{\tilde{\zeta}^{\mathrm{D}}_{j} \gamma^{\mathrm{D}}_{j}}{1+\gamma^{\mathrm{D}}_{j}} + \sum_{k=1}^{K} \frac{\tilde{\zeta}^{\mathrm{C}}_{k} \gamma^{\mathrm{C}}_{k}}{1+\gamma^{\mathrm{C}}_{k}} \nonumber  \\
\mathrm{s.t.} \quad
& \text{(\ref{eq:gc_cond})}, \ \text{(\ref{eq:phase_cond})}, \nonumber
\end{align}
where $\tilde{\zeta}^{\mathrm{D}}_{j}=1+\zeta^{\mathrm{D}}_{j}$ and $\tilde{\zeta}^{\mathrm{C}}_{k}=1+\zeta^{\mathrm{C}}_{k}$.
Plugging (\ref{eq:th_gd_re}) and (\ref{eq:th_gc_re}) into $\tilde{\digamma}(\bm{\phi})$, we have
\begin{align}
\tilde{\digamma}(\bm{\phi}) &= \sum_{j=1}^{J} \frac{ \tilde{\zeta}^{\mathrm{D}}_{j} A^{d}_{j}(\bm{\phi}) }{ A^{d}_{j}(\bm{\phi}) + \sum_{k=1}^{K} \rho_{jk} B^{c}_{kj}(\bm{\phi}) + \sigma_d^2} \nonumber\\
&+ \sum_{k=1}^{K} \frac{ \tilde{\zeta}^{\mathrm{C}}_{k} A^{c}_{k}(\bm{\phi}) }{ A^{c}_{k}(\bm{\phi}) + \sum_{j=1}^{J} \rho_{jk} B^{d}_{j}(\bm{\phi}) + \sigma_b^2}.
\label{eq:dual_tr}
\end{align}
To solve the MRFP problem (\ref{eq:dual_tr}), we apply the quadratic transform \cite{8314727} to (\ref{eq:dual_tr}). Then $\tilde{\digamma}(\bm{\phi})$ can be rewritten as (\ref{eq:Fq}) at the top of the next page,
\begin{figure*}[t]
\begin{align}
\tilde{\digamma}_{\mathrm{q}}(\bm{\phi},\bm{\xi}_{\mathrm{D}},\bm{\xi}_{\mathrm{C}}) &= \sum_{j=1}^{J} 2\sqrt{\tilde{\zeta}^{\mathrm{D}}_{j}} \mathrm{Re}\left\{{\xi^{\mathrm{D}}_{j}}^{\ast} ( \tilde{g}^{\mathrm{D}}_{j} + \mathbf{a}_{j}^{\mathsf H}\bm{\phi}  )\right\}
- \vert\xi^{\mathrm{D}}_{j}\vert^2 \Big( A^{d}_{j}(\bm{\phi}) + \sum_{k=1}^{K} \rho_{jk} B^{c}_{kj}(\bm{\phi}) + \sigma_d^2 \Big)
\nonumber\\
+& \sum_{k=1}^{K} 2\sqrt{\tilde{\zeta}^{\mathrm{C}}_{k}} \mathrm{Re}\left\{{\xi^{\mathrm{C}}_{k}}^{\ast} ( \tilde{g}^{\mathrm{C}}_{k} + \bm{\alpha}_{k}^{\mathsf H}\bm{\phi} )\right\}
- \vert\xi^{\mathrm{C}}_{k}\vert^2 \Big( A^{c}_{k}(\bm{\phi}) + \sum_{j=1}^{J} \rho_{jk} B^{d}_{j}(\bm{\phi}) + \sigma_b^2 \Big). \label{eq:Fq}
\end{align}
\hrulefill
\end{figure*}
where $\bm{\xi}_{\mathrm{D}}=[\xi^{\mathrm{D}}_{1},\cdots,\xi^{\mathrm{D}}_{J}]^{\mathsf{T}}$ and $\bm{\xi}_{\mathrm{C}}=[\xi^{\mathrm{C}}_{1},\cdots,\xi^{\mathrm{C}}_{K}]^{\mathsf{T}}$ are auxiliary variables introduced by quadratic transform. We optimize $\{\bm{\xi}_{\mathrm{D}},\bm{\xi}_{\mathrm{C}}\}$ and $\bm\phi$ in an alternating manner. The optimal $\{\xi^{\mathrm{D}}_{j},\xi^{\mathrm{C}}_{k}\}$ under a given $\bm\phi$ can be computed by setting their first derivatives to zero, as given by
\begin{align}
\hat{\xi}^{\mathrm{D}}_{j} &= \frac{\sqrt{\tilde{\zeta}_{\mathrm{D},j}} \left( \tilde{g}^{\mathrm{D}}_{j} + \mathbf{a}_{j}^{\mathsf H}\bm{\phi} \right) }{A^{d}_{j}(\bm{\phi}) + \sum_{k=1}^{K} \rho_{jk} B^{c}_{kj}(\bm{\phi}) + \sigma_d^2}, \label{eq:opt_xd}
\\
\hat{\xi}^{\mathrm{C}}_{k} &= \frac{\sqrt{\tilde{\zeta}_{\mathrm{C},k}} \left( \tilde{g}^{\mathrm{C}}_{k} + \bm{\alpha}_{k}^{\mathsf H}\bm{\phi} \right) }{A^{c}_{k}(\bm{\phi}) + \sum_{j=1}^{J} \rho_{jk} B^{d}_{j}(\bm{\phi}) + \sigma_b^2}.
\label{eq:opt_xc}
\end{align}
We proceed to optimize $\bm{\phi}$, given $\{\xi^{\mathrm{D}}_{j},\xi^{\mathrm{C}}_{k}\}$. Expanding the squared terms in (\ref{eq:Fq}), we have
\begin{align}
\tilde{\digamma}_{\mathrm{q}}(\bm{\phi}) = - \bm{\phi}^{\mathsf{H}} \mathbf{\Upsilon} \bm{\phi} + 2 \mathrm{Re}\{ \mathbf{u}^{\mathsf{H}} \bm{\phi}\} + C, \label{eq:qcqp}
\end{align}
where $\mathbf{\Upsilon}$, $\mathbf u$ and $C$ are given in (\ref{eq:long_1})-(\ref{eq:long_3}) at the bottom of the next page.
\begin{figure*}[b]
\hrulefill
\begin{align}
\mathbf{\Upsilon} &= \sum_{j=1}^{J} \vert\xi^{\mathrm{D}}_{j}\vert^2 \Big( \mathbf{a}_{j}\mathbf{a}_{j}^{\mathsf H} + \sum_{k=1}^{K} \rho_{jk} \mathbf{b}_{kj}\mathbf{b}_{kj}^{\mathsf{H}} \Big) + \sum_{k=1}^{K} \vert\xi^{\mathrm{C}}_{k}\vert^2 \Big( \bm{\alpha}_{k}\bm{\alpha}_{k}^{\mathsf{H}} + \sum_{j=1}^{J} \rho_{jk} \bm{\beta}_{j}\bm{\beta}_{j}^{\mathsf{H}} \Big), \label{eq:long_1}\\
\mathbf{u} &= \sum_{j=1}^{J} \sqrt{\tilde{\zeta}^{\mathrm{D}}_{j}} \xi^{\mathrm{D}}_{j} \mathbf{a}_{j} - \vert\xi^{\mathrm{D}}_{j}\vert^2 \Big( \tilde{g}^{\mathrm{D}}_{j} \mathbf{a}_{j} + \sum_{k=1}^{K} \rho_{jk} \tilde{f}^{\mathrm{C}}_{kj}\mathbf{b}_{kj} \Big) + \sum_{k=1}^{K} \sqrt{\tilde{\zeta}^{\mathrm{C}}_{k}} \xi^{\mathrm{C}}_{k} \bm{\alpha}_{k} - \vert\xi^{\mathrm{C}}_{k}\vert^2 \Big( \tilde{g}^{\mathrm{C}}_{k}\bm{\alpha}_{k} + \sum_{j=1}^{J} \rho_{jk} \tilde{f}^{\mathrm{D}}_{j}\bm{\beta}_{j} \Big), \label{eq:long_2} \\
{C} &= \sum_{j=1}^{J} 2\sqrt{\tilde{\zeta}^{\mathrm{D}}_{j}} \mathrm{Re}\{{\xi^{\mathrm{D}}_{j}}^{\ast} \tilde{g}^{\mathrm{D}}_{j} \} - \vert\xi^{\mathrm{D}}_{j}\vert^2 \Big( \vert \tilde{g}^{\mathrm{D}}_{j} \vert^2  + \sum_{k=1}^{K} \rho_{jk} \vert \tilde{f}^{\mathrm{C}}_{kj} \vert^2 + \sigma_d^2 \Big) \nonumber \\
&\qquad\qquad\qquad\qquad\qquad\qquad + \sum_{k=1}^{K} 2\sqrt{\tilde{\zeta}^{\mathrm{C}}_{k}} \mathrm{Re}\{ {\xi^{\mathrm{C}}_{k}}^{\ast} \tilde{g}^{\mathrm{C}}_{k} \} - \vert\xi^{\mathrm{C}}_{k}\vert^2 \Big( \vert \tilde{g}^{\mathrm{C}}_{k} \vert^2  + \sum_{j=1}^{J} \rho_{jk} \vert \tilde{f}^{\mathrm{D}}_{j}\vert^2 + \sigma_b^2 \Big). \label{eq:long_3}
\end{align}
\end{figure*}
After dropping the constant terms in (\ref{eq:qcqp}) and expanding the squared terms in constraint (\ref{eq:gc_cond}), Problem (P6) can be reformulated as
\begin{align}
\text{(P7)}: & \quad  \underset{\bm{\phi}}{\max} \quad -\bm{\phi}^{\mathsf{H}} \mathbf{\Upsilon} \bm{\phi} + 2 \mathrm{Re}\{ \mathbf{u}^{\mathsf{H}} \bm{\phi}\} \nonumber  \\
\mathrm{s.t.} & \quad \bm{\phi}^{\mathsf{H}} \mathbf{\Upsilon}^{\mathrm{C}}_{k} \bm{\phi} - 2 \mathrm{Re}\{ \mathbf{v}_{k}^{\mathsf{H}} \bm{\phi} \} \le \delta_{k},  \forall k, \label{eq:qcqp_c}\\
& \quad \text{(\ref{eq:phase_cond})}, \nonumber
\end{align}
where $\mathbf{\Upsilon}^{\mathrm{C}}_{k}=\gamma_{\mathrm{C}}^{\mathrm{th}}\sum_{j=1}^{J}\rho_{jk}\bm{\beta}_{j}\bm{\beta}_{j}^{\mathsf{H}}-\bm{\alpha}_{k}\bm{\alpha}_{k}^{\mathsf{H}}$, $\mathbf{v}_{k}=\tilde{g}^{\mathrm{C}}_{k}\bm{\alpha}_{k}-\gamma_{\mathrm{C}}^{\mathrm{th}}\sum_{j=1}^{J}\rho_{jk} \tilde{f}^{\mathrm{D}}_{j}\bm{\beta}_{j}$, and $\delta_{k}=\vert \tilde{g}^{\mathrm{C}}_{k} \vert^2-\gamma_{\mathrm{C}}^{\mathrm{th}} (\sigma_b^2 + \sum_{j=1}^{J}\rho_{jk}\vert \tilde{f}^{\mathrm{D}}_{j} \vert^2)$.

We note that Problem (P7) is an inhomogeneous QCQP problem \cite{5447068}, where constraint (\ref{eq:phase_cond}) is a non-convex unit-modulus constraint.
A popular method for solving (P7) is Gaussian randomization-based SDR \cite{8811733}, which relaxes the QCQP to a semidefinite program (SDP). In this case, the number of variables grows quadratically. The computational complexity is $\mathcal{O}((NL)^6)$ \cite{8982186}. Moreover, the randomization in SDR cannot guarantee a rank-one solution.

\section{Alternative Algorithm for Problem (P7)}\label{section:props}

\subsection{Consensus-ADMM Framework for Passive Beamforming}
We propose the new RM-ADMM by first introducing auxiliary variables $\{\mathbf{z}_k \}_{k=1}^{K}$ associated with the QCQP constraints (\ref{eq:qcqp_c}) and then transforming (P7) to the following consensus form:
\begin{subequations}
\begin{align}
\text{(P8)}: & \quad  \underset{\bm{\phi}\in\mathcal{M}, \{\mathbf{z}_k\}}{\min} \quad f(\bm{\phi}) = \bm{\phi}^{\mathsf{H}} \mathbf{\Upsilon} \bm{\phi} - 2 \mathrm{Re}\{ \mathbf{u}^{\mathsf H} \bm{\phi}\} \nonumber  \\
\mathrm{s.t.} & \quad \mathbf{z}_k^{\mathsf H} \mathbf{\Upsilon}^{\mathrm{C}}_{k} \mathbf{z}_k - 2 \mathrm{Re}\{ \mathbf{v}_{k}^{\mathsf H} \mathbf{z}_k \} \le \delta_{k},  \forall k, \label{eq:inequ_c} \\
& \quad \mathbf{z}_k = \bm{\phi}, \forall k, \label{eq:equ_c}
\end{align}
\end{subequations}
where the Riemannian submanifold $\mathcal{M}=\{\bm{\phi} \in \mathbb{C}^{NL}:\vert \theta_{l,n} \vert = 1\}$ is formed by the unit-modulus constraint (\ref{eq:phase_cond}).
For Problem (P8), the scaled form of ADMM \cite{Boyd101561} is given by
\begin{align}
\bm{\phi} &\leftarrow \underset{\bm{\phi}\in\mathcal{M}}{\mathrm{argmin}} \ \tilde{f}(\bm{\phi}) =f(\bm{\phi}) + \rho \sum_{k=1}^{K} \Vert \mathbf{z}_k-\bm{\phi}+\mathbf{r}_k \Vert^2, \label{eq:up_1} \\
\mathbf{z}_k &\leftarrow \underset{\mathbf{z}_k}{\mathrm{argmin}} \ \Vert \mathbf{z}_k-\bm{\phi}+\mathbf{r}_k \Vert^2,  \quad
\mathrm{s.t.} \ \text{(\ref{eq:inequ_c})}, \label{eq:up_2}\\
\mathbf{r}_k &\leftarrow   \mathbf{r}_k + \mathbf{z}_k - \bm{\phi}, \label{eq:up_3}
\end{align}
where $\rho$ is the penalty parameter and $\mathbf{r}_k$ is the scaled dual variable associated with the inequality constraint (\ref{eq:equ_c}).

Since the update in (\ref{eq:up_3}) is straightforward, we focus on subproblems (\ref{eq:up_1}) and (\ref{eq:up_2}).
The update in (\ref{eq:up_1}) can be handled using the standard Riemannian gradient descent (RGD) method. More details of RGD can be found in \cite{8125771}. To adopt the RGD, we first compute the Euclidean gradient of $\tilde{f}(\bm{\phi})$ at $\bm{\phi}_i$, i.e.,  $\Delta_{\bm{\phi}}\tilde{f}=2(\mathbf{\Upsilon}+K\rho\mathbf{I})\bm{\phi}_i-2[\mathbf{u}+\rho\sum_{k=1}^{K}(\mathbf{z}_k+\mathbf{r}_k)]$. The corresponding Riemannian gradient is $\Delta_{\mathcal{M}}\tilde{f}=\Delta_{\bm{\phi}}\tilde{f}-\mathrm{Re}\{\Delta_{\bm{\phi}} \circ \bm{\phi}_i^{\ast}\}\circ \bm{\phi}_i$ via the projection operator. The descent is performed
with a step $\alpha$ to arrive at the point $\bm{\phi}_i-\alpha \Delta_{\mathcal{M}}\tilde{f}$. To map the point $\bm{\phi}_i-\alpha \Delta_{\mathcal{M}}\tilde{f}$ back to $\mathcal{M}$, a retraction operator is performed to obtain $\bm{\phi}_{i+1}=\frac{\bm{\phi}_i-\alpha \Delta_{\mathcal{M}}\tilde{f}}{\Vert \bm{\phi}_i-\alpha \Delta_{\mathcal{M}}\tilde{f}\Vert}$.

\begin{figure*}[t]
	\centering{}\includegraphics[scale=0.4]{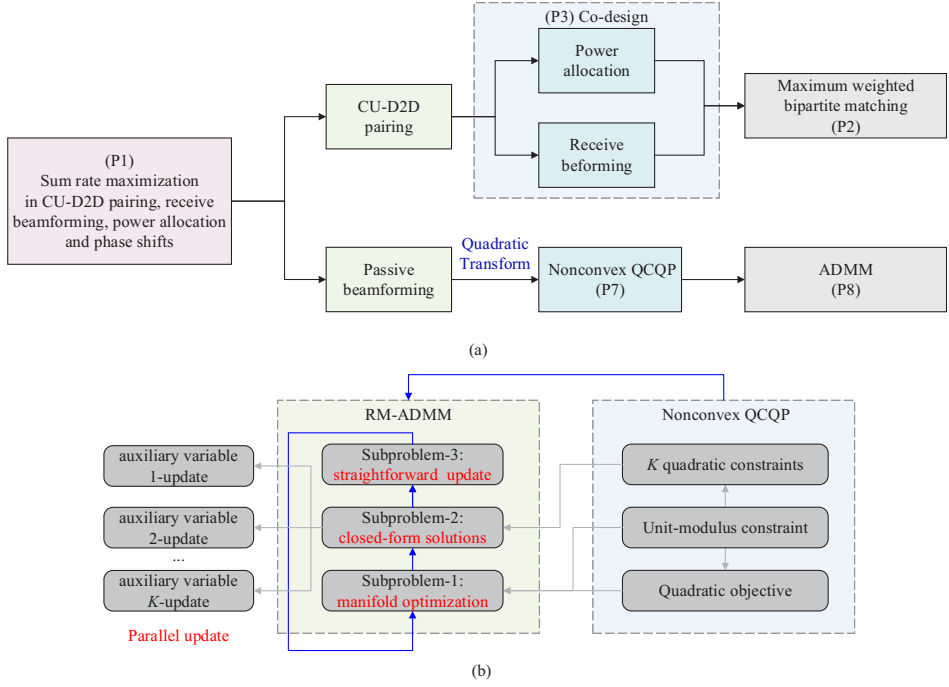}
	\caption{The flow diagram of the proposed algorithm. (a) Overall BCD algorithm framework; (b) RM-ADMM algorithm.}
	\label{fig:flow}
\end{figure*}

As per the update in (\ref{eq:up_2}), the proposed RM-ADMM results in $K$ simple QCQPs with only a single constraint each.
We first check whether $\bm{\phi}-\mathbf{r}_k$ satisfies constraint (\ref{eq:inequ_c}). If $\bm{\phi}-\mathbf{r}_k$ satisfies constraint (\ref{eq:inequ_c}), the solution of $\mathbf{z}_k$ is $\bm{\phi}-\mathbf{r}_k$. Otherwise, the optimal point of $\mathbf{z}_k$ can only be taken when constraint (\ref{eq:inequ_c}) holds with equality, according to the complementary slackness.
By setting the gradient of the Lagrangian of (\ref{eq:up_2}) to zero, we have
\begin{align}
\mathbf{z}_k (\mu) = (\mathbf{I}_{NL}+\mu \mathbf{\Upsilon}^{\mathrm{C}}_{k})^{-1}(\bm{\phi}-\mathbf{r}_k + \mu \mathbf{v}_{k}),
\label{eq:opt_z}
\end{align}
where $\mu$ is a Lagrange multiplier. By substituting (\ref{eq:opt_z}) into constraint (\ref{eq:inequ_c}) and setting the constraint with equality, then we have
\begin{equation}
g(\mu) = \mathbf{z}_k^{\mathsf{H}}(\mu) \mathbf{\Upsilon}^{\mathrm{C}}_{k} \mathbf{z}_k(\mu) - 2 \mathrm{Re}\{ \mathbf{v}_{k}^{\mathsf{H}} \mathbf{z}_k(\mu) \} - \delta_{k}.
\label{eq:bisec}
\end{equation}
By further defining $\tilde{\mathbf{r}}_k = \bm{\phi}-\mathbf{r}_k$, we can rewrite (\ref{eq:bisec}) as a nonlinear equation with respect to $\mu$:
\begin{equation}
g(\mu) = \sum_{i=1}^{NL} \epsilon_i \left\vert \frac{\tilde{{r}}_{ki} + \mu {v}_{ki}}{1 + \mu\epsilon_i} \right\vert^2 - 2\mathrm{Re} \left\{ \sum_{i=1}^{NL} {v}_{ki}^{\ast}  \frac{\tilde{{r}}_{ki} + \mu {v}_{ki}}{1 + \mu\epsilon_i}  \right\} - \delta_{k},
\label{eq:bisec2}
\end{equation}
where $\tilde{{r}}_{ki}$ and ${v}_{ki}$ are the $i$-th element of vectors $\tilde{\mathbf{r}}_k$ and $\mathbf{v}_{ki}$, respectively; and $\epsilon_1 \le \epsilon_2 \le \cdots \le \epsilon_{NL}$ are the eigenvalues of $\mathbf{\Upsilon}^{\mathrm{C}}_{k}$.
By taking the derivative of (\ref{eq:bisec2}), we have
\begin{align}
g{'}(\mu) = -2\sum_{i=1}^{NL} \frac{\vert {v}_{ki}-\epsilon_i\tilde{{r}}_{ki} \vert^2}{(1+\mu \epsilon_i)^3} < 0.
\end{align}
Hence, $g(\mu)$ is a monotonically decreasing function of $\mu$. The optimal $\mu$ can be obtained by bisection search. Let $\epsilon_{\max}$ and $\epsilon_{\min}$ be the maximum and minimum eigenvalues of $\mathbf{\Upsilon}^{\mathrm{C}}_{k}$, respectively. It can be shown that $g(-1/\epsilon_{\max})=+\infty$ and $g(-1/\epsilon_{\min})=-\infty$. Therefore, the initial interval of the bisection search can be specified as $(-1/\epsilon_{\max}, -1/\epsilon_{\min})$.

With the scaled form of the ADMM, the original intractable QCQP problem is converted into three sequential subproblems, i.e., (23)--(25),  and the unit-modulus constraints are eliminated in multiple quadratic constraints. As a result, we can invoke the manifold optimization to solve (23), and derive closed-form solutions for (24) that can be evaluated in a decentralized fashion. (25) only involves linear operations, and thus can be updated straightforwardly.
A series of new steps developed in the proposed BCD algorithm are depicted in Fig. \ref{fig:flow}.

\begin{algorithm}[htbp]
\small
\caption{RM-ADMM-based BCD Algorithm}
\label{alg:bcd}
\LinesNumbered
\SetKwInOut{KwIn}{\textbf{Initialize}}
\KwIn{$\{ \bm{\rho}, \mathbf{p}^{\mathrm{C}}, \mathbf{p}^{\mathrm{D}}, \mathbf{W}, \mathcal{S}_{\Theta} \}$.}
\Repeat{The objective value of (P1) converges}{
      \SetKwBlock{Begin}{Step 1:}{end}
      \Begin(\textbf{D2D-CU pairing}){
      Given $\bm{\phi}$, for all $j$ and $k$, compute $\{\mathbf{w}_k, P^{\mathrm{C}}_{k}, P^{\mathrm{D}}_{j}\}$ according to (6) and Proposition 1\;
      Given $\bm{\phi}$ and $\{\mathbf{w}_k, P^{\mathrm{C}}_{k}, P^{\mathrm{D}}_{j}\}$, calculate $R_{j,k}$ for all $j$ and $k$ to construct the BG for D2D-CU pairing\;
      Solve the D2D-CU pairing to update $\bm{\rho}$ by using the Hungarian algorithm\;}
      \SetKwBlock{Begin}{Step 2:}{end}
      \Begin(\textbf{Passive beamforming}){
      Given $\{P^{\mathrm{C}}_{k}, P^{\mathrm{D}}_{j}, \mathbf{w}_k, \bm{\phi}, \bm{\rho}\}$, calculate $\zeta^{\mathrm{D}}_{j}=\gamma^{\mathrm{D}}_{j}$ and $\zeta^{\mathrm{C}}_{k}=\gamma^{\mathrm{C}}_{k}$ by using (10) and (11)\;
      Given $\{P^{\mathrm{C}}_{k}, P^{\mathrm{D}}_{j}, \mathbf{w}_k, \bm{\phi}, \bm{\rho}, \zeta^{\mathrm{D}}_{j},\zeta^{\mathrm{C}}_{k}\}$, update $\{\xi^{\mathrm{D}}_{j},\xi^{\mathrm{C}}_{k}\}$ by using (15) and (16)\;
      Initialize $\{\mathbf{z}_k, \mathbf{r}_k\}$\;
      \Repeat{The objective value of (P7) converges}{
      Given $\{P^{\mathrm{C}}_{k}, P^{\mathrm{D}}_{j}, \mathbf{w}_k, \bm{\rho}, \zeta^{\mathrm{D}}_{j},\zeta^{\mathrm{C}}_{k}, \xi^{\mathrm{D}}_{j},\xi^{\mathrm{C}}_{k}, \mathbf{z}_k, \mathbf{r}_k\}$, update $\bm{\phi}$ by solving Problem (23) with the RGD method\;
      \eIf{$\bm{\phi}-\mathbf{r}_{k}$ feasible for (22a)}
      {$\mathbf{z}_{k} \leftarrow \bm{\phi}-\mathbf{r}_{k}$\;}
      {Given $\{P^{\mathrm{C}}_{k}, P^{\mathrm{D}}_{j}, \mathbf{w}_k, \bm{\rho}, \zeta^{\mathrm{D}}_{j},\zeta^{\mathrm{C}}_{k}, \xi^{\mathrm{D}}_{j},\xi^{\mathrm{C}}_{k}, \bm{\phi}, \mathbf{r}_k\}$, update $\mathbf{z}_{k}$ by using (26)\;}
      Given $\{P^{\mathrm{C}}_{k}, P^{\mathrm{D}}_{j}, \mathbf{w}_k, \bm{\rho}, \zeta^{\mathrm{D}}_{j},\zeta^{\mathrm{C}}_{k}, \xi^{\mathrm{D}}_{j},\xi^{\mathrm{C}}_{k}, \mathbf{z}_k, \bm{\phi}\}$, update $\mathbf{r}_{k}$ by using (25)\;
      }
      }
    }
\end{algorithm}

\subsection{Convergence and Complexity}

\emph{Convergence:} The above two-step alternating optimization chain comprises the overall BCD framework, which is described in Algorithm 1.
By following the BCD principle, Algorithm 1 increases iteratively the sum-rate of the considered system by updating the power allocation of the users, D2D-CU pairing, the receive beamformer of the BS, and the passive beamforming of the RISs, in an alternating manner until convergence. Specifically, for any configuration of the RISs $\{\theta_l\}_{l=1}^L$ (i.e., $\bm\phi$), the optimal power allocation of the users $\{P_j^D, P_k^C\}$ and receive beamforming vector of the BS $\mathbf{w}_k$ are obtained in closed-form.
With the closed-form expressions for $\{\mathbf{w}_k, P_{j}^{\mathrm{D}}, P_{k}^{\mathrm{C}}\}$ when given $\bm\phi$, the D2D-CU pairing $\bm{\rho}$ is interpreted as a maximum weighted BG matching problem and solved using the Hungarian algorithm in Step 1.
Given $\{P^{\mathrm{C}}_{k}, P^{\mathrm{D}}_{j}, \mathbf{w}_k, \bm{\phi}, \bm{\rho}\}$, the objective value of Problem (P7) is further increased by optimizing the phase shifts of the RISs $\{\theta_l\}_{l=1}^L$ with the proposed RM-ADMM algorithm in Step 2.
By running these two steps in an alternating fashion, the objective of Problem (P7) increases monotonically. On the other hand, the sum-rate of the considered system cannot grow unboundedly, given the power limits of the users.
As a result, the convergence of Algorithm 1 is guaranteed.

\emph{Complexity:}
The proposed RM-ADMM algorithm offers parallelizability with a lower computational complexity, as compared with the traditional SDR algorithms.
In the proposed RM-ADMM algorithm, the computational complexity of the RGD method is $\mathcal{O}((NL)^{2})$ in each iteration, and the computational complexity of the matrix inversion (\ref{eq:opt_z}) is $\mathcal{O}((NL)^{3})$.
In contrast, the SDR-based algorithms have a much higher computational complexity of $\mathcal{O}((NL)^6)$, which is prohibitive when $NL$ is large. Moreover, the large number of the QCQP constraints regarding $\bm{\phi}$ in Problem (P7) are decoupled in the proposed RM-ADMM algorithm. In other words,  we can efficiently update each $\mathbf{z}_k$ in parallel. The proposed RM-ADMM algorithm also provides closed-form solution for each independent QCQP constraint.

Problem (P7) is a nonconvex inhomogeneous QCQP with unit-modulus complex-valued variables, and solved using the proposed ADMM framework.
By using the SDR, the problem can be transformed to an SDP problem which could also be solved using the ADMM framework. However, the transformation from QCQP to SDP would increase quadratically the number of variables and, in turn, raises the computational complexity. Moreover, the use of ADMM to solve SDP problems also requires additional steps to recover $\bm{\phi}$. To this end, it is more efficient to solve the QCQP problem (P7) directly using the ADMM framework, as compared to solving the SDP problem transformed from Problem (P7).

\subsection{Extension Under Imperfect CSI}

Consider the widely adopted statistical CSI error model \cite{zhao2020exploiting}. The channel from the $k$-th CU to the BS is $\mathbf g_{k}^{\mathrm{C}}=\hat{\mathbf g}_{k}^{\mathrm{C}} + \Delta \mathbf g_{k}^{\mathrm{C}}$, where $\mathbf g_{k}^{\mathrm{C}}$ is the actual CSI, $\hat{\mathbf{g}}_k^{\mathrm{C}}$ is the estimated CSI, and $\Delta\mathbf{g}_k^{\mathrm{C}}$ is the channel estimation error with elements following the circularly symmetric complex Gaussian (CSCG) distribution, i.e., $\Delta \mathbf{g}_k^{\mathrm{C}} \sim \mathcal{CN}(0, \epsilon_{g,C,k}^2)$.
Likewise, the channel from the $j$-th D2D transmitter (DT) to the $j$-th D2D receiver (DR) is $g_{j}^{\mathrm{D}}=\hat{g}_{j}^{\mathrm{D}}+\Delta g_{j}^{\mathrm{D}}$.
The interference channels from the $k$-th CU to the $j$-th DR and from the $j$-th DT to the BS are $f_{kj}^{\mathrm{C}}=\hat{f}_{kj}^{\mathrm{C}} + \Delta f_{kj}^{\mathrm{C}}$ and $\mathbf f_{j}^{\mathrm{D}}=\hat{\mathbf f}_{j}^{\mathrm{D}}+\Delta \mathbf f_{j}^{\mathrm{D}}$, respectively.
The cascaded channel from the $j$-th DT to the $l$-th RIS and then to the $j$-th DR is denoted by
${\mathbf{s}^{\mathrm{r}}_{lj}}^{\mathsf H} \mathsf{Diag}\{\mathbf{s}^{\mathrm{t}}_{lj}\} \triangleq \mathbf{q}_{1,lj}^{\mathsf H} = \hat{\mathbf{q}}_{1,lj}^{\mathsf H} + \Delta \mathbf{q}_{1,lj}^{\mathsf H}$.
Then, the cascaded channel from the $k$-th CU to the $l$-th RIS and then to the $j$-th DR is ${\mathbf{s}^{\mathrm{r}}_{lk}}^{\mathsf{H}}\mathsf{Diag}\{\mathbf{s}^{\mathrm{C}}_{lk}\} \triangleq \mathbf{q}_{2,lkj}^{\mathsf H} = \hat{\mathbf{q}}_{2,lkj}^{\mathsf H} + \Delta \mathbf{q}_{2,lkj}^{\mathsf H}$.
The cascaded channel from the $k$-th CU to the $l$-th RIS and then to the BS is
$\mathbf{S}^{\mathrm{B}}_{l} \mathsf{Diag}\{\mathbf{s}^{\mathrm{C}}_{lk}\} \triangleq \mathbf{Q}_{1,lk} = \hat{\mathbf{Q}}_{1,lk} + \Delta \mathbf{Q}_{1,lk}$.
The cascaded channel from the $j$-th DT to the $l$-th RIS and then to the BS is
$\mathbf{S}^{\mathrm{B}}_{l} \mathsf{Diag}\{\mathbf{s}^{\mathrm{t}}_{lj} \} \triangleq \mathbf{Q}_{2,lj} = \hat{\mathbf{Q}}_{2,lj} + \Delta \mathbf{Q}_{2,lj}$.

Under the statistical CSI error model, the SINR at the $j$-th DR is written as
\small
\begin{align}
\gamma^{\mathrm{D}}_{j}
&= \frac{ {P^{\mathrm{D}}_{j}} \vert \hat{g}^{\mathrm{D}}_{j} + \sum_{l=1}^{L} \hat{\mathbf{q}}_{1,lj}^{\mathsf H} \bm{\theta}_l \vert^2}{ {\Delta}_{1,j} +  {\sum_{k=1}^{K}\rho_{jk} P^{\mathrm{C}}_{k}} \vert \hat{f}^{\mathrm{C}}_{kj} + \sum_{l=1}^{L} \hat{\mathbf{q}}_{2,lkj}^{\mathsf H} \bm{\theta}_l\vert^2 + \sigma_{d}^2},
\end{align}\normalsize
and the SINR of the $k$-th CU at the BS is written as
\small
\begin{align}
\gamma^{\mathrm{C}}_{k}
&= \frac{ {P^{\mathrm{C}}_{k}}\vert \mathbf{w}_{k}^{\mathsf{H}} ( \hat{\mathbf{g}}^{\mathrm{C}}_{k}+ \sum_{l=1}^{L} \hat{\mathbf{Q}}_{1,lk} \bm{\theta}_l  ) \vert^2}{ {\Delta}_{2,k} + \sum_{j=1}^{J}\rho_{jk} {P^{\mathrm{D}}_{j}}\vert \mathbf{w}_{k}^{\mathsf{H}} ( \hat{\mathbf{f}}^{\mathrm{D}}_{j} + \sum_{l=1}^{L} \hat{\mathbf{Q}}_{2,lj} \bm{\theta}_l) \vert^2 + \sigma_b^2},
\end{align}\normalsize
where the interference terms resulting from imperfect CSI, i.e., ${\Delta}_{1,j}$ and ${\Delta}_{2,k}$, are given by
\begin{align}
{\Delta}_{1,j} &= {P^{\mathrm{D}}_{j}} \vert \Delta{g}^{\mathrm{D}}_{j} + \sum_{l=1}^{L} \Delta{\mathbf{q}}_{1,lj}^{\mathsf H} \bm{\theta}_l \vert^2
\nonumber\\
& \quad +
{\sum_{k=1}^{K}\rho_{jk} P^{\mathrm{C}}_{k}} \vert \Delta{f}^{\mathrm{C}}_{kj} + \sum_{l=1}^{L} \Delta{\mathbf{q}}_{2,lkj}^{\mathsf H} \bm{\theta}_l\vert^2, \\
{\Delta}_{2,k} &= {P^{\mathrm{C}}_{k}}\vert \mathbf{w}_{k}^{\mathsf{H}} ( \Delta{\mathbf{g}}^{\mathrm{C}}_{k}+ \sum_{l=1}^{L} \Delta{\mathbf{Q}}_{1,lk} \bm{\theta}_l  ) \vert^2 \nonumber\\
& \quad + \sum_{j=1}^{J}\rho_{jk} {P^{\mathrm{D}}_{j}}\vert \mathbf{w}_{k}^{\mathsf{H}} ( \Delta{\mathbf{f}}^{\mathrm{D}}_{j} + \sum_{l=1}^{L} \Delta{\mathbf{Q}}_{2,lj} \bm{\theta}_l) \vert^2.
\end{align}
After substituting (30) and (31) into problem (P1), the problem becomes intractable due to the lack of a closed-form sum-rate in the objective of the problem.
We can resort to maximizing tractable lower bound for the expected achievable sum-rate.
Based on \cite[Proposition~1]{zhao2020exploiting}, the transmit rate of the $j$-th D2D link, and the transmit rate of the $k$-th CU are respectively lower-bounded by
\small
\begin{align}
&\mathbb{E}(R_{j}^{\mathrm{D}}) \ge \widetilde{R}_{j}^{\mathrm{D}} \nonumber\\
=& \log \left( 1+ \frac{ {P^{\mathrm{D}}_{j}} \vert \hat{g}^{\mathrm{D}}_{j} + \sum_{l=1}^{L} \hat{\mathbf{q}}_{1,lj}^{\mathsf H} \bm{\theta}_l \vert^2}{ {\sum_{k=1}^{K}\rho_{jk} P^{\mathrm{C}}_{k}} \vert \hat{f}^{\mathrm{C}}_{kj} + \sum_{l=1}^{L} \hat{\mathbf{q}}_{2,lkj}^{\mathsf H} \bm{\theta}_l\vert^2 + \mathbb{E}({\Delta}_{1,j}) +\sigma_{d}^2}\right) \nonumber\\
=& \log \left( 1+ \frac{ {P^{\mathrm{D}}_{j}} \vert \hat{g}^{\mathrm{D}}_{j} + \hat{\mathbf{q}}_{1,j}^{\mathsf H} \bm{\phi} \vert^2}{ {\sum_{k=1}^{K}\rho_{jk} P^{\mathrm{C}}_{k}} \vert \hat{f}^{\mathrm{C}}_{kj} + \hat{\mathbf{q}}_{2,kj}^{\mathsf H} \bm{\phi} \vert^2 + \mathbb{E}({\Delta}_{1,j}) +\sigma_{d}^2}\right) ,  \label{eq:e_d}\\
&\mathbb{E}(R_{k}^{\mathrm{C}}) \ge \widetilde{R}_{k}^{\mathrm{C}} \nonumber\\
=& \log \left( 1+ \frac{ {P^{\mathrm{C}}_{k}}\vert \mathbf{w}_{k}^{\mathsf{H}} ( \hat{\mathbf{g}}^{\mathrm{C}}_{k}+ \sum_{l=1}^{L} \hat{\mathbf{Q}}_{1,lk} \bm{\theta}_l  ) \vert^2}{  \sum_{j=1}^{J}\rho_{jk} {P^{\mathrm{D}}_{j}}\vert \mathbf{w}_{k}^{\mathsf{H}} ( \hat{\mathbf{f}}^{\mathrm{D}}_{j} + \sum_{l=1}^{L} \hat{\mathbf{Q}}_{2,lj} \bm{\theta}_l) \vert^2 + \mathbb{E}({\Delta}_{2,k}) + \sigma_b^2} \right) \nonumber\\
=& \log \left( 1+ \frac{ {P^{\mathrm{C}}_{k}}\vert \mathbf{w}_{k}^{\mathsf{H}} ( \hat{\mathbf{g}}^{\mathrm{C}}_{k}+  \hat{\mathbf{Q}}_{1,k} \bm{\phi}  ) \vert^2}{  \sum_{j=1}^{J}\rho_{jk} {P^{\mathrm{D}}_{j}}\vert \mathbf{w}_{k}^{\mathsf{H}} ( \hat{\mathbf{f}}^{\mathrm{D}}_{j} + \hat{\mathbf{Q}}_{2,j} \bm{\phi} ) \vert^2 + \mathbb{E}({\Delta}_{2,k}) + \sigma_b^2} \right),\label{eq:e_c}
\end{align}\normalsize
where $\hat{\mathbf{q}}_{1,j}=\mathsf{vec}([\hat{\mathbf{q}}_{1,1j},\cdots, \hat{\mathbf{q}}_{1,Lj}])$, $\hat{\mathbf{q}}_{2,kj}=\mathsf{vec}([\hat{\mathbf{q}}_{2,1kj}, \cdots, \hat{\mathbf{q}}_{2,Lkj}])$, $\hat{\mathbf{Q}}_{1,k} = [\hat{\mathbf{Q}}_{1,1k}, \cdots, \hat{\mathbf{Q}}_{1,Lk}]$, $\hat{\mathbf{Q}}_{2,j}=[\hat{\mathbf{Q}}_{2,1j}, \cdots, \hat{\mathbf{Q}}_{2,Lj}]$, and $\bm{\phi}=\mathsf{vec}([\bm{\theta}_1,\cdots,\bm{\theta}_L])$. $\mathbb{E}({\Delta}_{1,j})$ and $\mathbb{E}({\Delta}_{2,k})$ are given by
\small
\begin{align}
\mathbb{E}({\Delta}_{1,j})&=  {P^{\mathrm{D}}_{j}} ( \epsilon_{g,D,j} + \epsilon_{q,1,j} \Vert \bm{\phi} \Vert^2) \nonumber\\
&+ {\sum_{k=1}^{K}\rho_{jk} P^{\mathrm{C}}_{k}} ( \epsilon_{f,C,kj} + \epsilon_{q,2,kj} \Vert \bm{\phi} \Vert^2), \\
\mathbb{E}({\Delta}_{2,k})&={P^{\mathrm{C}}_{k}} (\epsilon_{g,C,k} + \epsilon_{Q,1,k} \Vert\bm{\phi} \Vert^2) \Vert \mathbf{w}_{k}\Vert^2 \nonumber\\
&+  \sum_{j=1}^{J}\rho_{jk} {P^{\mathrm{D}}_{j}} (\epsilon_{f,D,j} + \epsilon_{Q,2,j} \Vert\bm{\phi} \Vert^2) \Vert \mathbf{w}_{k} \Vert^2\nonumber\\
&= \mathcal{E}_{2,k} \mathbf{w}_{k}^{\mathsf H} \mathbf{w}_{k}.
\end{align}\normalsize
Here, $\mathcal{E}_{2,k} = {P^{\mathrm{C}}_{k}} (\epsilon_{g,C,k} + \epsilon_{Q,1,k} \Vert\bm{\phi} \Vert^2) + \sum_{j=1}^{J}\rho_{jk} {P^{\mathrm{D}}_{j}} (\epsilon_{f,D,j} + \epsilon_{Q,2,j} \Vert\bm{\phi} \Vert^2)$.
Now, Problem (P1) is replaced by the maximization of the lower bound of the expected achievable sum-rate, as given by
\begin{align}
\text{(P9)}:  &\ \underset{\{\bm{\rho}, \mathbf{p}^{\mathrm{C}}, \mathbf{p}^{\mathrm{D}}, \mathbf{w}, \mathcal{S}_{\Theta}\}}{\max} \ \sum_{k=1}^{K} \widetilde{R}_{k}^{\mathrm{C}} +  \sum_{j=1}^{J} \widetilde{R}_{j}^{\mathrm{D}} \nonumber  \\
\mathrm{s.t.}  &  \quad \widetilde{R}_{k}^{\mathrm{C}} \ge \log( 1+\gamma_{\mathrm{C}}^{\mathrm{th}}), \ \forall k\in \mathcal{C},\\
\quad &  \quad \text{(\ref{eq:pd})}, \text{(\ref{eq:pc})}, \text{(\ref{eq:re_bf})}-\text{(\ref{eq:phase})},\nonumber
\end{align}
First, we tackle the constraint (38) by maximizing $\widetilde{R}_{k}^{\mathrm{C}}$ to obtain an optimal receive beamformer $\mathbf{w}$.
Plugging (37) into (35), we can solve the receive beamforming subproblem, i.e.,
\begin{align}
\mathbf{w} (P^{\mathrm{D}}) &= \underset{\mathbf{w}: \|\mathbf{w}^{2}\|{=}1}{\arg\max} \quad \frac {P_{C}\mathbf{w}^{\mathsf H}\hat{\mathbf{h}}^{\mathrm{C}} \hat{\mathbf{h}}^{\mathrm{C}^{\mathsf H}}\mathbf{w}}{\mathbf{w}^{\mathsf H} \left( P^{\mathrm{D}} \hat{\mathbf{h}}^{\mathrm{D}} \hat{\mathbf{h}}^{{\mathrm{D}}^{\mathsf H}}  + (\mathcal{E}_{2,k} + \sigma_b^2) \mathbf{I}_M \right)\mathbf{w}}
\nonumber\\
&= \frac{  [P^{\mathrm{D}} \hat{\mathbf{h}}^{\mathrm{D}} \hat{\mathbf{h}}^{{\mathrm{D}}^{\mathsf H}}  + (\mathcal{E}_{2,k} + \sigma_b^2) \mathbf{I}_M ]^{-1} \hat{\mathbf{h}}^{\mathrm{C}} }{\Vert (P^{\mathrm{D}} \hat{\mathbf{h}}^{\mathrm{D}} \hat{\mathbf{h}}^{{\mathrm{D}}^{\mathsf H}}  + (\mathcal{E}_{2,k} + \sigma_b^2) \mathbf{I}_M )^{-1} \hat{\mathbf{h}}^{\mathrm{C}} \Vert}.
\end{align}
Note that $\mathbb{E}({\Delta}_{1,j})$ and $\mathbb{E}({\Delta}_{2,k})$ are constant in (\ref{eq:e_d}) and (\ref{eq:e_c}), given fixed $\mathbf{w}$ and $\bm{\phi}$.
Therefore, we can rewrite (\ref{eq:e_d}) and (\ref{eq:e_c}) as
\small
\begin{align}
\widetilde{R}_{j}^{\mathrm{D}}
&= \log \left( 1+ \frac{ {P^{\mathrm{D}}_{j}} \vert \hat{g}^{\mathrm{D}}_{j} + \hat{\mathbf{q}}_{1,j}^{\mathsf H} \bm{\phi} \vert^2}{ {\sum_{k=1}^{K}\rho_{jk} P^{\mathrm{C}}_{k}} \vert \hat{f}^{\mathrm{C}}_{kj} + \hat{\mathbf{q}}_{2,kj}^{\mathsf H} \bm{\phi} \vert^2 + \sigma_{1,j}^2}\right),  \\
\widetilde{R}_{k}^{\mathrm{C}}
&= \log \left( 1+ \frac{ {P^{\mathrm{C}}_{k}}\vert \mathbf{w}_{k}^{\mathsf{H}} ( \hat{\mathbf{g}}^{\mathrm{C}}_{k}+  \hat{\mathbf{Q}}_{1,k} \bm{\phi}  ) \vert^2}{  \sum_{j=1}^{J}\rho_{jk} {P^{\mathrm{D}}_{j}}\vert \mathbf{w}_{k}^{\mathsf{H}} ( \hat{\mathbf{f}}^{\mathrm{D}}_{j} + \hat{\mathbf{Q}}_{2,lj} \bm{\phi} ) \vert^2 +  \sigma_{2,k}^2} \right),
\end{align}\normalsize
where $\sigma_{1,j}^2=\mathbb{E}({\Delta}_{1,j}) +\sigma_{d}^2$ and $\sigma_{2,k}^2=\mathbb{E}({\Delta}_{2,k}) + \sigma_b^2$.
It can be found that the optimal power solution $(\hat{P}^{\mathrm{D}}_j, \hat{P}^{\mathrm{C}}_k)$ can still be obtained based on Proposition 1, after replacing $\tilde{\gamma}_{\mathrm{C}}=\frac{\sigma_b^2 \gamma_{\mathrm{C}}^{\mathrm{th}}}{\Vert \mathbf{h}^{\mathrm{C}} \Vert^2}$ and $\lambda_{2}=\frac{\sigma_b^2}{\Vert \mathbf{h}^{\mathrm{D}} \Vert^2}$ in Proposition 1 with $\tilde{\gamma}_{\mathrm{C}}=\frac{\sigma_2^2 \gamma_{\mathrm{C}}^{\mathrm{th}}}{\Vert \mathbf{h}^{\mathrm{C}} \Vert^2}$ and $\lambda_{2}=\frac{\sigma_2^2}{\Vert \mathbf{h}^{\mathrm{D}} \Vert^2}$, respectively.

After constructing the cost matrix of $\widetilde{R}_{j,k}=\widetilde{R}_{j}^{\mathrm{D}} + \widetilde{R}_{k}^{\mathrm{C}}$, the D2D-CU matching
aims to maximize the expected achievable sum-rate, i.e., $\underset{  \bm{\rho}  }{\max} \  \sum_{k\in\mathcal{C}} \sum_{j\in\mathcal{D}} \rho_{jk} \widetilde{R}_{j}^{\mathrm{D}} +  \widetilde{R}_{k}^{\mathrm{C}}(P^{\mathrm{C}}_k, \rho_{jk} P^{\mathrm{D}}_j)$, which can still be interpreted as a maximum weighted bipartite matching problem. As a result, the Hungarian algorithm remains valid.

Given $\{ \bm\rho, \mathbf{p}^{\mathrm{D}}, \mathbf{p}^{\mathrm{C}} ,\mathbf{w}\}$, we can see the resulting passive beamforming subproblem from Problem (P9) is still a QCQP:
\begin{align}
\text{(P10)}: & \quad  \underset{\bm{\phi}}{\max} \quad -\bm{\phi}^{\mathsf{H}} \mathbf{\Upsilon} \bm{\phi} + 2 \mathrm{Re}\{ \mathbf{u}^{\mathsf{H}} \bm{\phi}\} \nonumber  \\
\mathrm{s.t.} & \quad \bm{\phi}^{\mathsf{H}} \mathbf{\Upsilon}^{\mathrm{C}}_{k} \bm{\phi} - 2 \mathrm{Re}\{ \mathbf{v}_{k}^{\mathsf{H}} \bm{\phi} \} \le \widetilde{\delta}_{k},  \forall k, \\
& \quad \text{(\ref{eq:phase_cond})}, \nonumber
\end{align}
where $\widetilde{\delta}_{k}=\vert \tilde{g}^{\mathrm{C}}_{k} \vert^2-\gamma_{\mathrm{C}}^{\mathrm{th}} (\sigma_{2,k}^2 + \sum_{j=1}^{J}\rho_{jk}\vert \tilde{f}^{\mathrm{D}}_{j} \vert^2)$.
Problem (P10) has the same structure as (P7), and can be readily solved by our proposed methods.

\section{Simulation Results}\label{section:sim}
Simulation results are provided to evaluate the performance of our proposed algorithm and the potential benefits of deploying the RISs in D2D systems.
The cell radius is 500 m. The CUs are uniformly distributed in the ring area situated between 400 m and 500 m from the BS.
The distance of D2D link is randomly and uniformly distributed in [10 m, 30 m] \cite{8927889}. In our simulations, $L=4$ RISs are placed at the cell edge. Their positions are (0, 500 m), (500 m, 0), (0, -500 m) and (-500 m, 0), respectively.
All the channels in the simulation consider the $L_0$-tap baseband equivalent multi-path channel \cite{9039554,9326394,7234857}.
In the absence of the RISs, the channel impulse response between the BS and the user (either a CU or DT) is given by
\begin{align}
\mathbf{g} (t) = \sum_{\ell=0}^{L_0-1} \alpha_{\ell} \bar{\mathbf{g}} (\ell) \delta(t-\tau_{\ell}),
\end{align}
where $L_0$ is the number of taps, $\alpha_{\ell}$ is the complex amplitude of the $\ell$-th tap, $\bar{\mathbf{g}}$ is the array steering vector for the angle-of-arrival (AoA) of the $\ell$-th tap, and $\delta(t)$ is the pulse shaping filter.
We assume that the number of taps $L_0$ is 16, the complex amplitude of each tap $\alpha_{\ell}$ follows the Rayleigh fading and the path loss exponent is 3.8 in the absence of the RISs.
Likewise, for the RIS-related channels (i.e., either RIS-BS or RIS-user link), we assume that the number of taps is 4, the complex amplitude of each tap follows the Rician fading with the Rician factor of 10 dB, and the path loss exponent of 2.2 \cite{8982186}.
For the channel between any two users (including CUs or DUs), we assume that the number of taps is 16, and the complex amplitude of each tap follows the Rayleigh fading, and path loss exponent is 4 in the absence of the RISs.
We set $M=4$, $\sigma_{d}^2=\sigma_{b}^2 = -115\ \text{dB}$, and $P_{\mathrm{C}}^{\max}=P_{\mathrm{D}}^{\max}=P$.

\begin{figure}[t]
	\centering
    \subfigure[]{\includegraphics[width=2.6in]{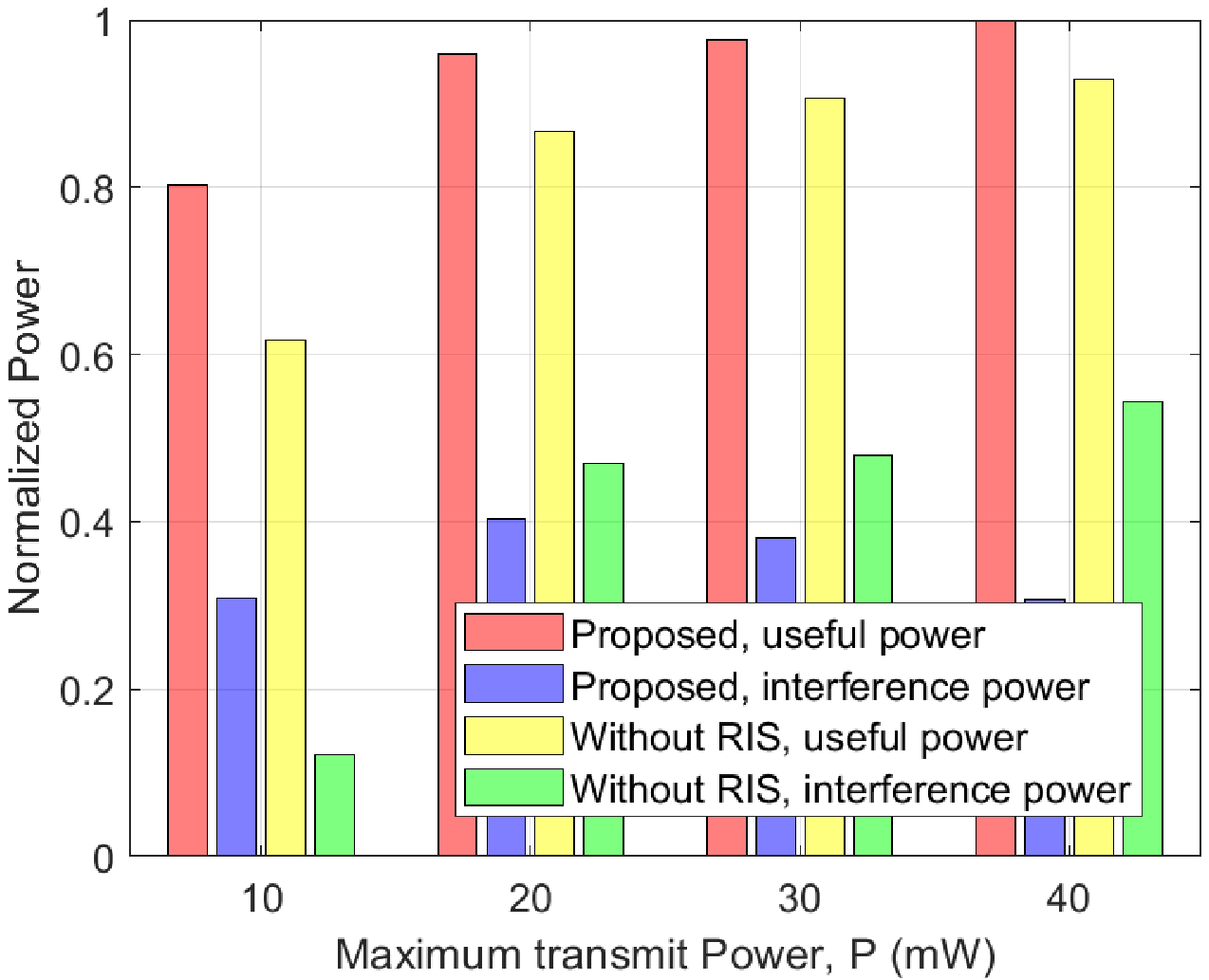}}
    \quad
    \subfigure[]{\includegraphics[width=2.6in]{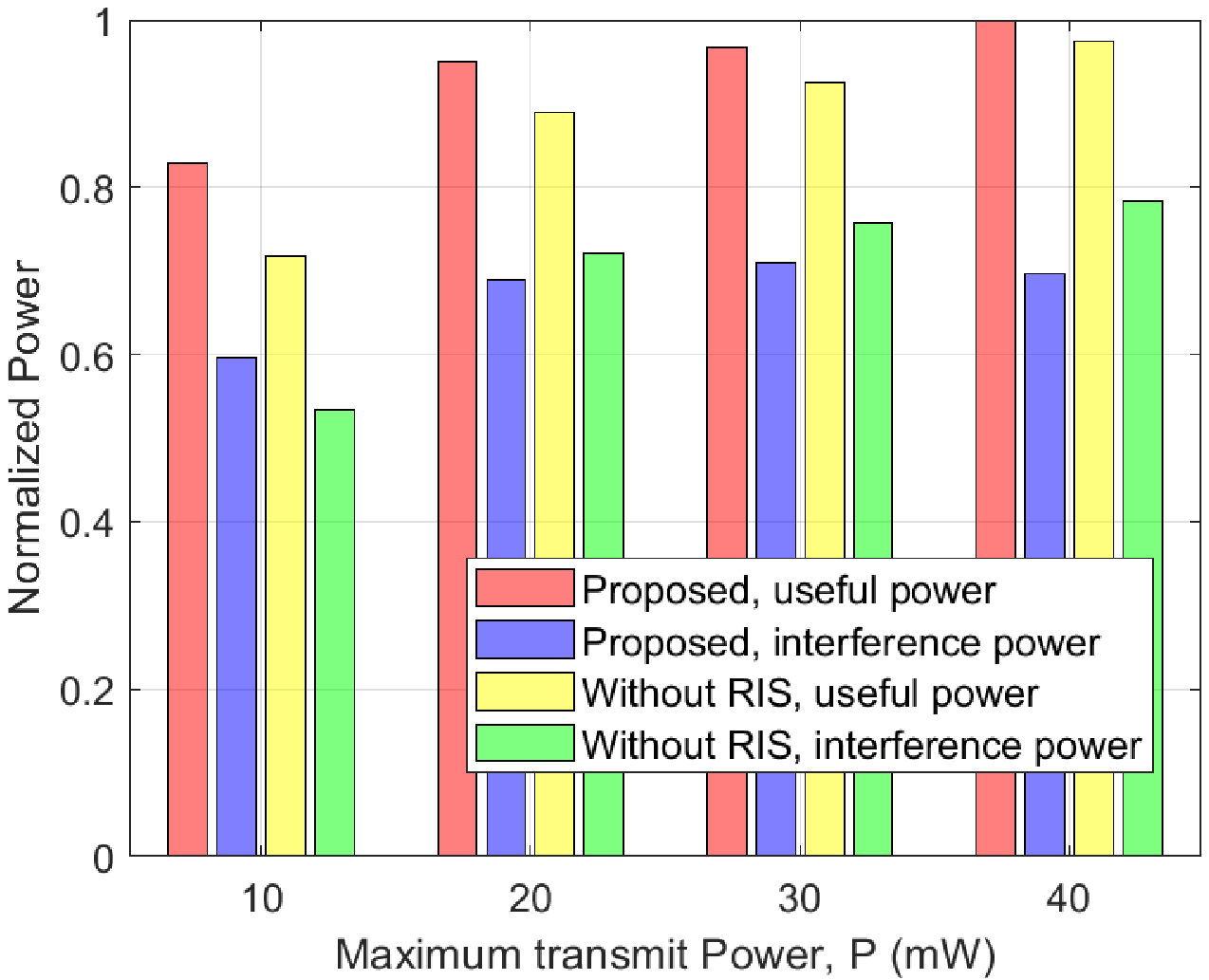}}
	\caption{Proportion of total power at the (a) BS; (b) DR.}
	\label{fig:p1}
\end{figure}

\subsection{Scenario of Single CU and D2D Pair}

\begin{figure}[t]
	\centering
    \subfigure[]{\includegraphics[width=2.6in]{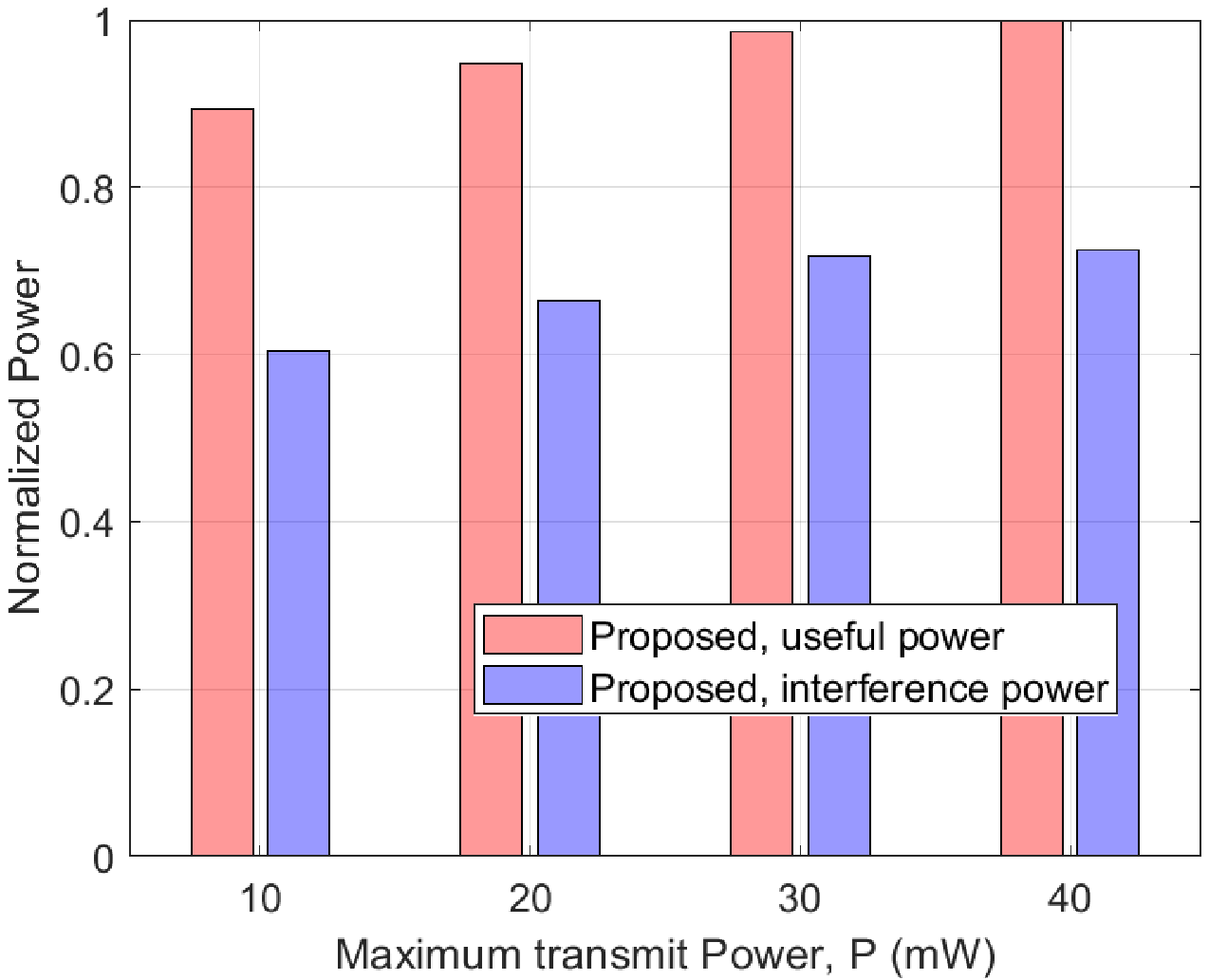}}
    \quad
    \subfigure[]{\includegraphics[width=2.6in]{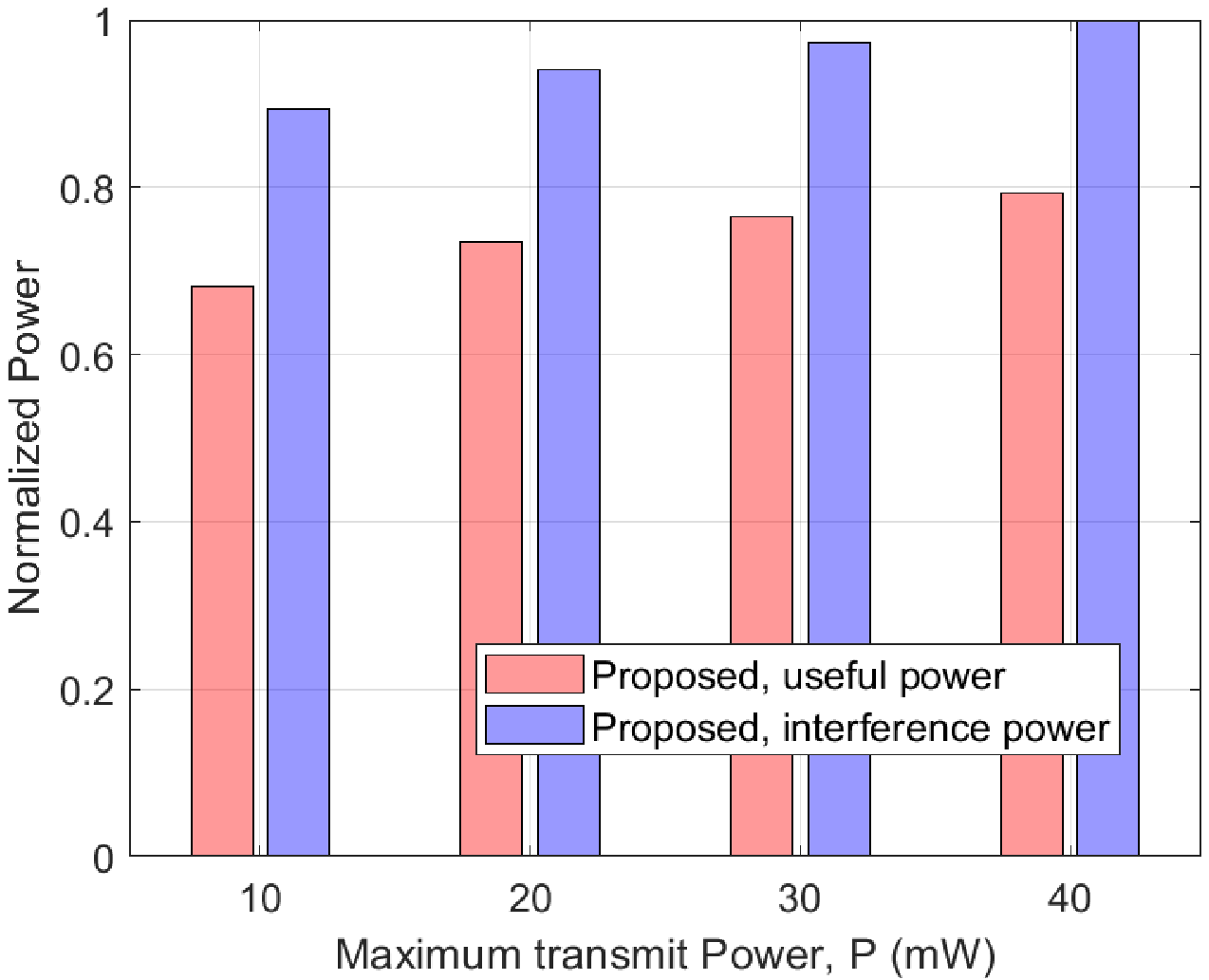}}
	\caption{Proportion of reflecting power at the (a) BS; (b) DR.}
	\label{fig:p2}
\end{figure}

We also provide the normalized signal/interference power versus the maximum transmit power to illustrate the role of RISs in the considered system.
The measured power values are normalized by the maximum power value of a set of powers.
Fig. \ref{fig:p1} shows the useful/interference power proportion in the total power (including the power from the direct links).
As shown in Figs. \ref{fig:p1}(a) and (b), the total interference powers grow at both the BS and DR, as the maximum power increases in the absence of the RISs.
In contrast, the total interference power is significantly reduced in the presence of the RISs.
Fig. 5 shows the useful/interference power proportion in the reflected power.
As shown in Fig. \ref{fig:p2}(a), the reflected useful power is always higher than the reflected interference power at the BS.
Conversely, as shown in Fig. \ref{fig:p2}(b), the gain of the reflected useful signals is lower than the gain of the reflected interference signals at the DRs.
This is reasonable, since higher gains of reflected useful signals are needed to improve the relatively weaker CU-BS links, as compared to the D2D links.

\begin{figure}[t]
    \centering{}\includegraphics[scale=0.5]{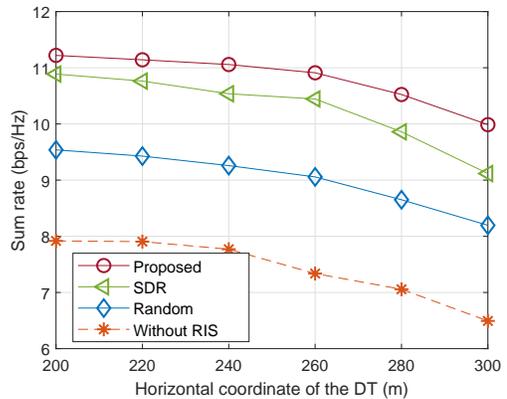}
	\caption{Achievable sum-rate versus horizontal coordinate of the DT.}
	\label{fig:0}
\end{figure}

\begin{figure}[t]
    \centering{}\includegraphics[scale=0.5]{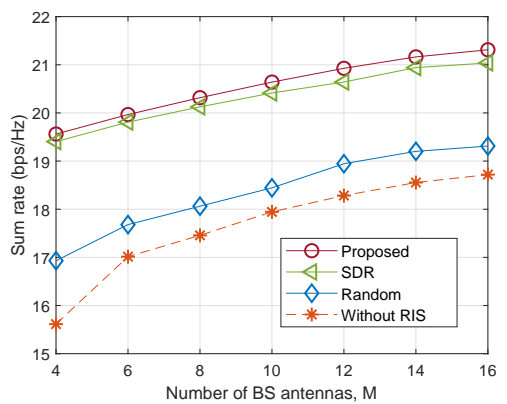}
	\caption{Achievable sum-rate versus the number of BS antennas.}
	\label{fig:1}
\end{figure}

In Fig. \ref{fig:0}, we investigate the performance of different schemes under different horizontal coordinate of the DT. One pair of D2D devices and one CU are considered. The CU is located at (400 m, 0) and the range of DT location is from (200 m, 0) to (300 m, 0). We set $P=20\ \text{mW}$ and $N=10$. We see that the achievable sum-rate decreases as the DT moves towards the CU. Since the distance from the DT to the CU is larger than the distance from the DT to the BS, the interference from CU to the D2D link becomes increasingly severe with the increasing horizontal coordinate of the DT.

Fig. \ref{fig:1} shows the achieved sum-rate of the considered approaches, under different numbers of BS antennas. We set $K=J=1$. The achievable sum-rate of each scheme grows, as the number of BS antennas increases. It can be concluded that increasing the number of antennas can enhance the receive beamforming gain and improve the system sum-rate.

\begin{figure}[t]
    \centering{}\includegraphics[scale=0.5]{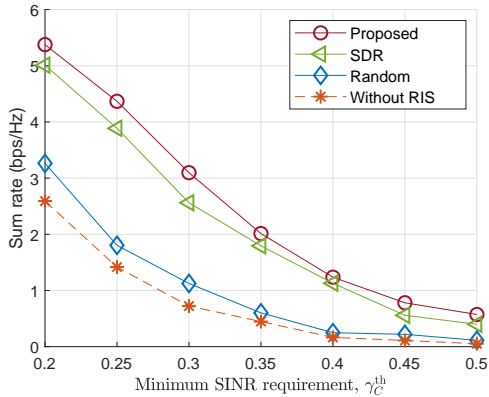}
	\caption{Achievable sum-rate versus the minimum SINR requirement of the CU.}
	\label{fig:2}
\end{figure}

Fig. \ref{fig:2} studies the sum-rates of the different schemes by varying the minimum SINR threshold of the CU $\gamma_{\mathrm{C}}^{\mathrm{th}}$. We set $K=J=1$, $P=20\ \text{mW}$, and $N=5$. The sum-rate of the proposed algorithm with RISs outperforms the other schemes when there are a single CU and a single D2D pair. We see that the achievable sum-rate decreases, as $\gamma_{\mathrm{C}}^{\mathrm{th}}$ increases. This is because the CU, located at the edge of the cell, is far away from the BS. The weak CU-BS channels can be compensated by reducing the transmit power of the DT and focusing the reflection gain of the RISs in the direction of the desired CU links. Therefore, the sum-rate of the DUs can be suppressed, despite the short distance between the DUs.

\begin{figure}[h]
	\centering{}\includegraphics[scale=0.5]{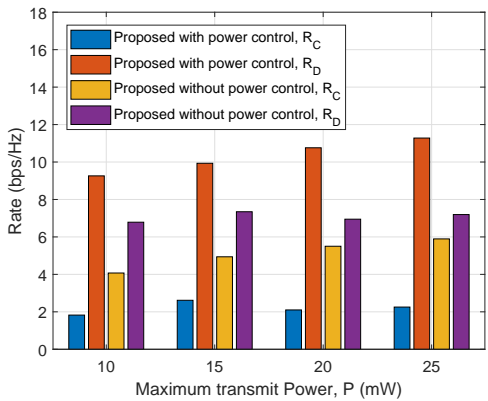}
	\caption{The proportion of the achievable rates of CU and D2D pair.}
	\label{fig:rate_prop}
\end{figure}

Fig. \ref{fig:rate_prop} shows the achievable data rates of the CUs and DUs with the growing maximum transmit power of the users.
For illustration convenience, we consider a single CU and a single D2D pair for illustration convenience, where the CU is at (400 m, 0), the DT is at (300 m, 0), and the DR is randomly distributed within 30 meters from the DT. For comparison purpose, we also consider the proposed BCD algorithm, when all the users persistently transmit the maximum power, referred to as ``Proposed without power control''.
It is observed that the data rate of the D2D pair is always higher than that of the CU, due to the fact that the D2D users are usually close to each other and enjoy good channel conditions.
When the users transmit the maximum power, the achievable data rate of the CU increases as the maximum transmit power increases, while the achievable rate of the D2D pair remains nearly unchanged.
This is because under the maximum transmit power, the interference from the D2D pair to the CU is high and the passive beamforming of the RISs is configured to enhance the cellular link. When the transmit powers are adaptively configured, the date rate of the CU can achieve its rate threshold while the data rate of the DU can increase to maximize the system rate.

\subsection{Scenario of Multiple CUs and D2D Pairs}

\begin{figure}[t]
	\centering{}\includegraphics[scale=0.5]{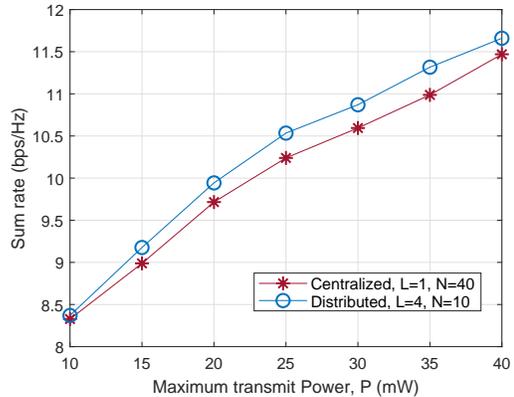}
	\caption{Comparison of sum-rate between centralized and distributed deployment of RISs.}
	\label{fig:case2}
\end{figure}

We study the impact of the RIS deployment on the sum-rate by considering the following multi-user setting shown in Fig. \ref{fig:case2}, where the centralized deployment places a single RIS with 40 passive elements at (500 m, 0), and the distributed deployment places 4 RISs with 10 passive elements per RIS at (0, 500 m), (500 m, 0), (0, -500 m) and (-500 m, 0).
We set $K=2$ and $J=2$.
Fig. \ref{fig:case2} shows that the sum-rate with four distributed, small RISs is higher than the sum-rate with the single large RIS.
One reason is that given the uniform distribution of the users around the cell edge, some users may be far away from both the BS and the single RIS under the centralized deployment of the single RIS.
Another reason is that the optimization of an RIS is increasingly constrained with the growing number of served users, due to the passive nature of the RIS. In other words, the gain of an RIS grows sublinearly with its size, given the number of users.

\begin{figure}[t]
    \centering{}\includegraphics[scale=0.5]{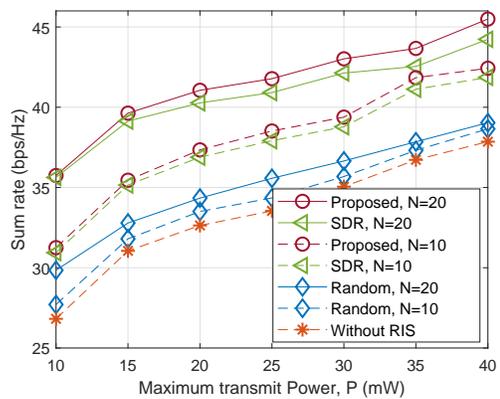}
	\caption{Achievable sum-rate versus maximum transmit power.}
	\label{fig:3}
\end{figure}

We also investigate the impact of the maximum transmit power on the sum-rate in Fig. \ref{fig:3}. We set $K=3$, $J=2$, and $\gamma_{\mathrm{C}}^{\mathrm{th}}=0.5$.
As expected, the sum-rate gains of all schemes improve significantly, as the maximum transmit power increases.
We observe that all schemes using RISs significantly outperform the scheme without the RIS.
By comparing the achievable sum-rate against the maximum transmit power under different numbers of RIS elements, we conclude that the large-scale passive RIS helps improve the achievable sum-rate, given the SINR requirements of the CUs.

\begin{figure}[t]
    \centering{}\includegraphics[scale=0.5]{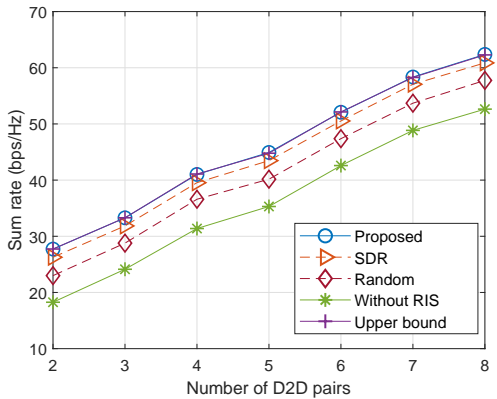}
	\caption{Achievable sum-rate with the number of D2D pairs.}
	\label{fig:4}
\end{figure}

\begin{figure}[t]
    \centering{}\includegraphics[scale=0.5]{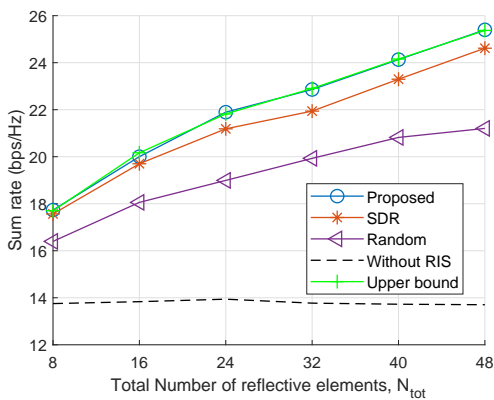}
	\caption{Achievable sum-rate with the number of total reflective elements.}
	\label{fig:5}
\end{figure}

\begin{figure}[t]
    \centering{}\includegraphics[scale=0.5]{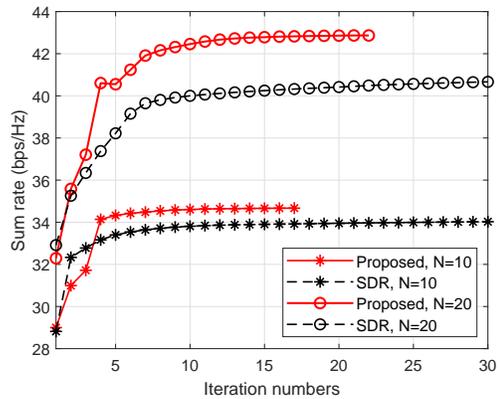}
	\caption{Convergence behavior of the proposed algorithm and SDR method.}
	\label{fig:6}
\end{figure}

Fig. \ref{fig:4} compares the achievable sum-rates of the considered schemes with a growing number of D2D pairs. We set $K=10$ and $\gamma_{\mathrm{C}}^{\mathrm{th}}=0.5$.
The upper bound of the sum-rate obtained by the relaxed Problem (P7) is plotted for comparison \cite{8723525}.
With the increasing number of D2D pairs, the sum-rate of the system grows under each scheme.
The achievable sum-rates of the schemes with RISs improve more significantly, as compared to those without RIS.
Moreover, the proposed algorithm is indistinguishably close to the upper bound, confirming the effectiveness of the algorithm.

Fig. \ref{fig:5} plots the achievable sum-rates versus the total number of reflective elements at the RISs.
As anticipated, the RIS with more reflective elements can offer a stronger beamforming gain while meeting the SINR requirements of both CU and D2D pairs.
Furthermore, the proposed algorithm is indistinguishably close to the upper bound of the achievable sum-rate with the increasing number of reflective elements, corroborating again the effectiveness of the algorithm.

Fig. \ref{fig:6} shows the convergence behavior of the proposed RM-ADMM-based BCD algorithm and the state-of-the-art SDR-based BCD method, where $K=3$, $J=2$, $P=20\ \text{mW}$, and $\gamma_{\mathrm{C}}^{\mathrm{th}}=0.5$. We can see that the sum-rate of the RM-ADMM-based BCD algorithm increases faster than that of the SDR-based BCD technique. The proposed algorithm only requires a few iterations to converge.
Moreover, the sum-rate gain of the proposed algorithm over the SDR-based BCD algorithm increases, as $N$ grows from
10 to 20.
The conclusion drawn is that the proposed algorithm is increasingly advantageous when the RISs are large.

\section{Conclusion}\label{section:con}
Considering the uplink of an RIS-assisted D2D-underlaid cellular system, we developed the new BCD algorithm to maximize the system sum-rate by jointly optimizing the power control, the D2D-CU pairing, the receive beamforming of the BS, and the passive beamforming of the RISs. An efficient RM-ADMM algorithm was proposed to solve the passive beamforming of the RISs, thus avoiding a prohibitive computational cost which would occur if the existing SDR-based techniques are applied.
Simulations showed that the RIS-assisted D2D-underlaid communication system can significantly improve the sum-rate, compared to the systems without the RISs. Our proposed algorithm can provide marked increases in sum-rate with a competitive computational complexity, as compared with its SDR-based alternative.

\begin{appendices}
\section{Proof of Proposition 1}\label{prf1}
Let $\nu_{0}=\vert h^{\mathrm{C}}_k \vert^2$, $\nu_{1}=\frac{\Vert \mathbf{h}^{\mathrm{C}}_k \Vert^2}{\sigma_b^2}$ and $\nu_{2}=\vert h^{\mathrm{D}}_j \vert^2$, we rewrite the objective of (P2) as
\begin{align}
R^{\mathrm{b}}(P^{\mathrm{C}}_k, P^{\mathrm{D}}_j) = &\log \Big[\big( 1 + \nu_{1} P^{\mathrm{C}}_k  \frac{\lambda_{2} + (1-\lambda_{1}) P^{\mathrm{D}}_j }{\lambda_{2}  + P^{\mathrm{D}}_j } \big) \nonumber\\
& \big(1 + \frac{\nu_{2} P^{\mathrm{D}}_j }{\nu_{0} P^{\mathrm{C}}_k + \sigma_d^2} \big) \Big]. \label{eq:sum_choose}
\end{align}
By invoking the boundary optimum existence lemma \cite{7933260}, we see that the optimal power pair lies on the vertical or horizontal border lines of $\mathcal{P}$.
Denote the vertical or horizontal border lines as $R^{\mathrm{C}}(P^{\mathrm{C}}_k)=R^{\mathrm{b}}(P_{\mathrm{D}}^{\max},P^{\mathrm{C}}_k)$ and $R^{\mathrm{D}}(P^{\mathrm{D}}_j)=R^{\mathrm{b}}(P^{\mathrm{D}}_j,P_{\mathrm{C}}^{\max})$, respectively.
Taking the first and second derivatives, we find that $R^{\mathrm{D}}(P^{\mathrm{D}}_j)$ is a strictly increasing function and $R^{\mathrm{C}}(P^{\mathrm{C}}_k)$ is either an increasing or convex function.
This proof completes. \QEDA

\section{Proof of Proposition 2}\label{prf2}
Note that $\digamma(\zeta,\gamma)=\log(1+\zeta)-\zeta+\frac{(1+\zeta)\gamma}{1+\gamma}$ is a concave and differentiable function of $\zeta$ when given $\gamma$. Thus, setting $\frac{\partial \digamma}{\partial \zeta}$ to zero yields $\hat{\zeta}=\gamma$. Based on this result, substituting the obtained solution $\{\hat{\zeta}^{\mathrm{D}}, \hat{\zeta}^{\mathrm{C}}\}$ into the objective of (P5) can lead to the objective function of (P4). In this sense, the optimal objective values of these two problems are equal. Their equivalence is established. \QEDA

\end{appendices}


\bibliographystyle{IEEEtran}

\begin{IEEEbiography}[{\includegraphics[width=1in,height=1.25in,clip,keepaspectratio]{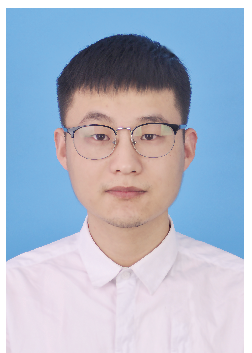}}]{Yashuai Cao}
(S'18) received the B.S. degree from Chongqing University of Posts and Telecommunications (CQUPT), Chongqing, China, in 2017. He is currently pursuing the Ph.D. degree in communication engineering with the School of Information and Communication Engineering, Beijing University of Posts and Telecommunications (BUPT), Beijing, China. His current research interests include wireless resource allocation and signal processing technologies for massive MIMO systems and intelligent reflecting surface assisted wireless networks.
\end{IEEEbiography}

\begin{IEEEbiography}[{\includegraphics[width=1in,height=1.25in,clip,keepaspectratio]{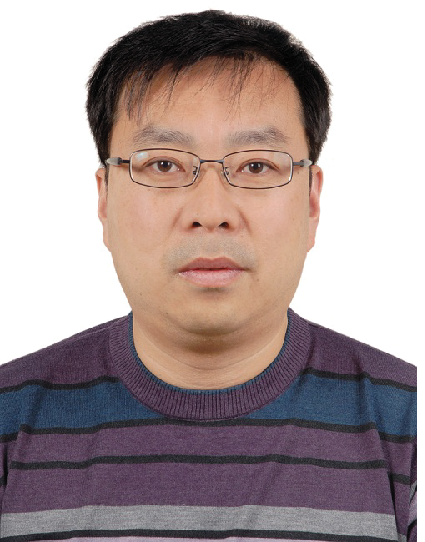}}]{Tiejun Lv}
(M'08-SM'12) received the M.S. and Ph.D. degrees in electronic engineering from the University of Electronic Science and Technology of China (UESTC), Chengdu, China, in 1997 and 2000, respectively. From January 2001 to January 2003, he was a Postdoctoral Fellow with Tsinghua University, Beijing, China. In 2005, he was promoted to a Full Professor with the School of Information and Communication Engineering, Beijing University of Posts and Telecommunications (BUPT). From September 2008 to March 2009, he was a Visiting Professor with the Department of Electrical Engineering, Stanford University, Stanford, CA, USA. He is the author of three books, more than 90 published IEEE journal papers and 200 conference papers on the physical layer of wireless mobile communications. His current research interests include signal processing, communications theory and networking. He was the recipient of the Program for New Century Excellent Talents in University Award from the Ministry of Education, China, in 2006. He received the Nature Science Award in the Ministry of Education of China for the hierarchical cooperative communication theory and technologies in 2015.
\end{IEEEbiography}

\begin{IEEEbiography}[{\includegraphics[width=1in,height=1.25in,clip,keepaspectratio]{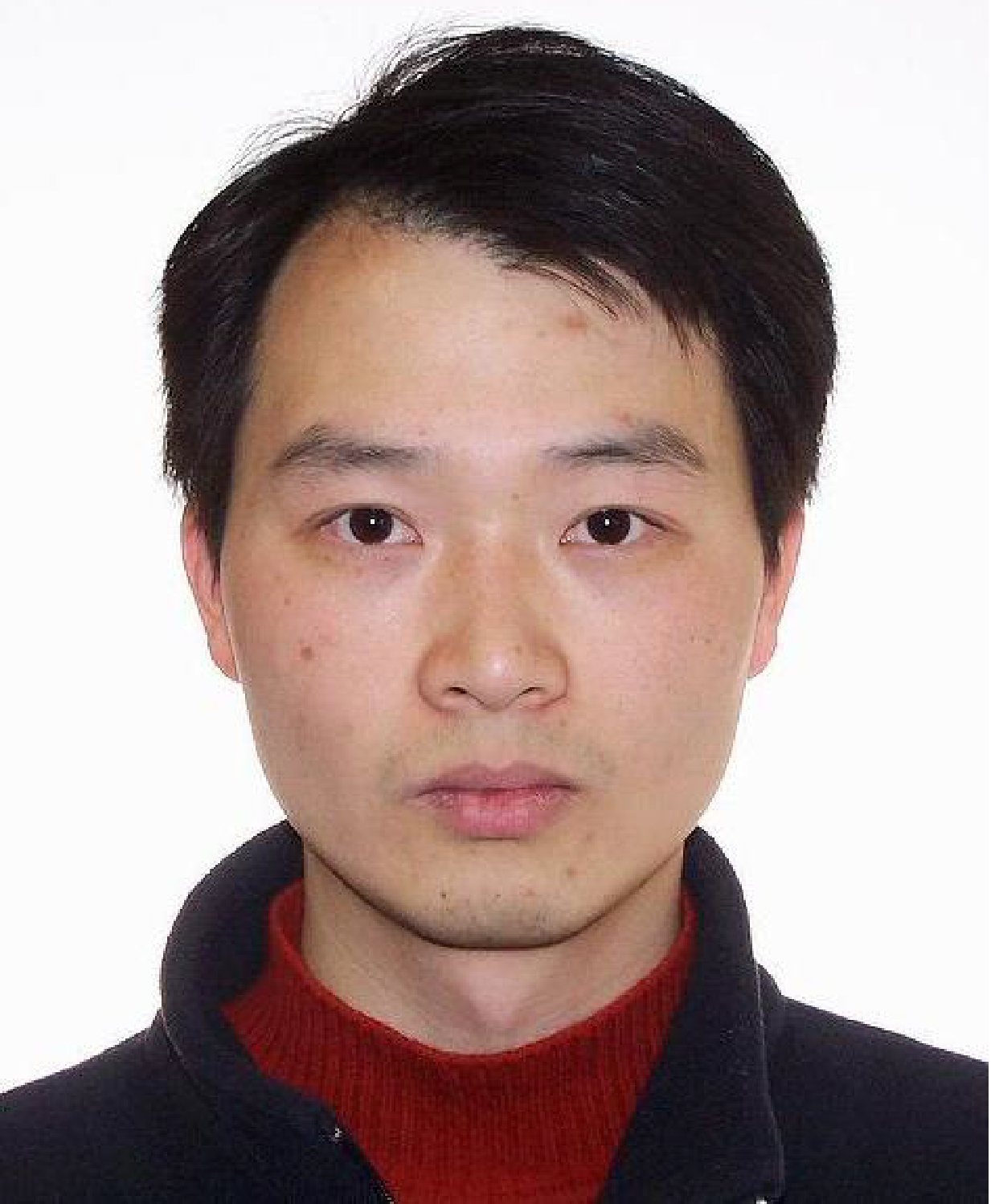}}]{Wei Ni}
(M'09-SM'15) received the B.E. and Ph.D. degrees in Electronic Engineering from Fudan University, Shanghai, China, in 2000 and 2005, respectively. Currently, he is a Group Leader and Principal Research Scientist at CSIRO, Sydney, Australia, and an Adjunct Professor at the University of Technology Sydney and Honorary Professor at Macquarie University, Sydney. He was a Postdoctoral Research Fellow at Shanghai Jiaotong University from 2005--2008; Deputy Project Manager at the Bell Labs, Alcatel/Alcatel-Lucent from 2005 to 2008; and Senior Researcher at Devices R\&D, Nokia from 2008 to 2009. His research interests include signal processing, stochastic optimization, learning, as well as their applications to network efficiency and integrity.\\
Dr. Ni is the Chair of IEEE Vehicular Technology Society (VTS) New South Wales (NSW) Chapter since 2020 and an Editor of IEEE Transactions on Wireless Communications since 2018. He served first the Secretary and then Vice-Chair of IEEE NSW VTS Chapter from 2015 to 2019, Track Chair for VTC-Spring 2017, Track Co-chair for IEEE VTC-Spring 2016, Publication Chair for BodyNet 2015, and Student Travel Grant Chair for WPMC 2014.
\end{IEEEbiography}

\begin{IEEEbiography}[{\includegraphics[width=1in,height=1.25in,clip,keepaspectratio]{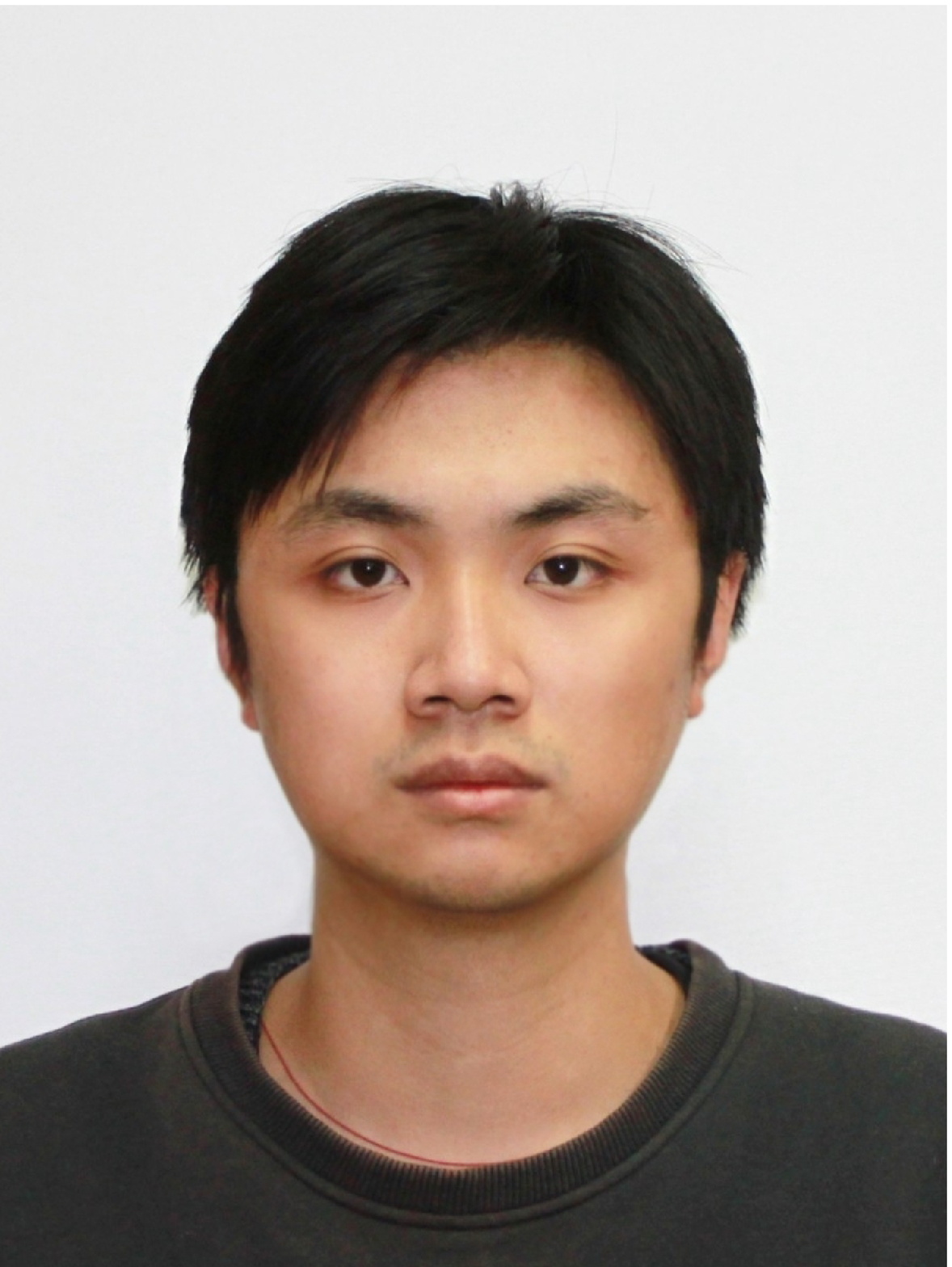}}]{Zhipeng Lin}
(M'20) received the Ph.D. degrees from the School of Information and Communication Engineering, Beijing University of Posts and Telecommunications, Beijing, China, and the School of Electrical and Data Engineering, University of Technology of Sydney, NSW, Australia, in 2021. Currently, He is an Associate Researcher in the College of Electronic and Information Engineering, Nanjing University of Aeronautics and Astronautics, Nanjing, China. His current research interests include signal processing, massive MIMO, hybrid beamforming, and UAV communications.
\end{IEEEbiography}

\end{document}